\documentclass[12pt]{article}

\begin{document}

\normalsize
\title{\normalsize\bf \boldmath $SU(3)$ - Flavor Symmetry in $B \rightarrow VP$ Decays \unboldmath}
\author{\normalsize Gil Paz \\ {\normalsize \it Physics Department, Technion-Israel Institute of Technology, 3200 Haifa, Israel}}
\date{}
\maketitle
\begin{abstract}
In the framework of $SU(3)$ symmetry, we present a general analysis of $B$ meson decays into two lighter uncharmed mesons (both pairs of pseudoscalar mesons and pairs of vector and pseudoscalar mesons). From the analysis we find constraints on $\gamma$ and discuss their validity. The most useful new constraint is obtained by considering the
decay modes $B^{0} \rightarrow K\pi$ and $B^{+} \rightarrow \pi^{0}\pi^{+}$.
In decays into pairs of vector and pseudeuscalar mesons, no constraints can be obtained using $SU(3)$ symmetry alone and further assumptions are needed. Based on these assumptions, we obtain new (weaker) constraints using $B^{0} \rightarrow \rho K$/$B^{0} \rightarrow K^{*}\pi$ and $B^{+} \rightarrow \rho\pi$. We show that no other constraints can be obtained. We also suggest a method to measure $\gamma$ using $B^{0}_{s} \rightarrow \rho\pi$ and $B^{0} \rightarrow K^{*\pm}K^{\mp}$. 
\end{abstract}
\leftline{PACS numbers: 13.25.Hw, 14.40.Nd, 11.30.Er}
\thispagestyle{empty}
\newpage
\section{Introduction}
The existence of CP violation in the standard model depends on one parameter that need to be different from zero, namely the complex phase of the CKM matrix. The existence of this phase is equivalent to a non zero area of the famous unitarity triangle ($V_{ud}V_{ub}^{*}+V_{cd}V_{cb}^{*}+V_{td}V_{tb}^{*}=0$).\\
The angle ``$\beta$'' of this triangle was recently measured from the time dependent $CP$ asymmetry of the decay $B \rightarrow J/\psi K_{S}$ \cite{beta}. The determination of the angle ``$\gamma$'' seem to be more difficult. One of the possible strategies is to use $SU(3)$ symmetry to measure or constrain $\gamma$ using $B$ mesons decays into two lighter uncharmed mesons\cite{grl,grlh,fleischer,nr,gpy,vp}\\
In general the decay modes receive contributions from ``Tree'', ``Penguin'' and ``ElectroWeak Penguin'' (EWP) amplitudes. Since the amplitudes involve several unknown parameters, it is hard, in general, to obtain useful data even if we decrease the number of the unknowns using $SU(3)$ symmetry. Neubert and Rosner \cite{nr} have suggested a way to relate the EWP contribution to the Tree contribution in the decay modes $B^{+} \rightarrow K^{+} \pi^{0}$ and $B^{+} \rightarrow K^{0} \pi^{+}$ and obtain a constraint on $\gamma$. Similar constraint were obtained later using other decay modes into pairs of pseudoscalar mesons ($PP$), namely $B^{0} \rightarrow K^{+} \pi^{-}$ and $B^{+} \rightarrow K^{0} \pi^{+}$ \cite{gpy}, and by using $B$ decays into pairs of vector and pseudoscalar mesons ($VP$), namely $B^{+} \rightarrow \rho K$ and $B^{+} \rightarrow K^{*} \pi$ \cite{vp}. The natural question arises, whether these are all the constraints that can be obtained. In order to answer this question, we will take a more systematic approach to the problem.\\
In section 2 we repeat the original Neubert-Rosner argument and identify its essential features. We then ``rearrange'' the effective Hamiltonian and use $SU(3)$ symmetry to write it in terms of irreducible operators. In this form the relations between the Tree, Penguin and EWP contributions will be more evident. Using Wigner-Eckart theorem we present (in appendix A) a decomposition of the various decay amplitudes (both $PP$ and $VP$). In section 3 we use this decomposition to obtain all the possible constraints and discuss their validity. As we shall see, new constraints can be obtained using $B^{0} \rightarrow K\pi$ and $B^{+} \rightarrow \pi^{0}\pi^{+}$ (for $PP$) and using $B^{0} \rightarrow \rho K$/$B^{0} \rightarrow K^{*}\pi$ and $B^{+} \rightarrow \rho\pi$ (for $VP$ with further assumptions). In section 4 we present our conclusions alongside a new method to measure $\gamma$\\

\section{Algebraic Analysis of the Decay Amplitudes}
\subsection{The Essential features of the Neubert-Rosner Method}
In order to obtain the essential features that are needed, let us repeat the Neubert-Rosner argument in the form formulated by Gronau, Pirjol and Yan \cite{gpy} . The suggestion of Neubert and Rosner is based on the decomposition of the following decay amplitudes \cite{gpy}:\\
\begin{eqnarray}
\sqrt{2}{\cal A}(B^{+} \rightarrow K^{+} \pi^{0}) & = &
-\lambda_{u}^{(s)} (T+C+P_{uc}+A) \\
&&-\lambda_{t}^{(s)} \left[ P_{ct}-\sqrt{2}P^{EW}(B^{+} \rightarrow K^{+} \pi^{0})\right]
 \nonumber \\
&& \nonumber \\ 
{\cal A}(B^{+} \rightarrow K^{0} \pi^{+}) & = &
+\lambda_{u}^{(s)}(P_{uc}+A) \\ 
&&+\lambda_{t}^{(s)} \left[P_{ct}-P^{EW}(B^{+} \rightarrow K^{0} \pi^{+})\right] 
  \nonumber\\ 
&& \nonumber \\
\label{measure} 
-\sqrt{2}{\cal A}(B^{+} \rightarrow \pi^{0} \pi^{+}) &= &
+\lambda_{u}^{(d)} (T+C), 
\end{eqnarray}
where $P^{EW}(\cdots)$ denotes the EWP contribution and $P_{uc}=P_{u}-P_{c}$,
 $P_{ct}=P_{c}-P_{t}$.\\
Adding up the first two expressions one gets:
\begin{eqnarray}
\sqrt{2}{\cal A}(B^{+} \rightarrow K^{+} \pi^{0})+
{\cal A}(B^{+} \rightarrow K^{0} \pi^{+})&=&
-\lambda_{u}^{(s)} (T+C) \\\
&& +\lambda_{t}^{(s)}\left[\sqrt{2}P^{EW}(B^{+} \rightarrow K^{+} \pi^{0})+
P^{EW}(B^{+} \rightarrow K^{0} \pi^{+})\right]\nonumber,
\end{eqnarray}
which resembles the amplitude of $B^{+} \rightarrow \pi^{0} \pi^{+}$ apart from the EWP contribution.\\
The EWP contribution can be related to $(T+C)$ \cite{gpy}:
\begin{equation}
\label{ewpt}
\sqrt{2}P^{EW}(B^{+} \rightarrow K^{+} \pi^{0})+
P^{EW}(B^{+} \rightarrow K^{0} \pi^{+})=\frac{3}{2}
\frac{c_{9}+c_{10}}{c_{1}+c_{2}}(T+C).
\end{equation}
Using the orthogonality relation\\
\begin{equation}
V_{us}V_{ub}^{*}+V_{cs}V_{cb}^{*}+V_{ts}V_{tb}^{*}=0,
\end{equation}
or
\begin{equation}
\label{ortho}
\lambda_{t}^{(s)}=-\lambda_{u}^{(s)}-\lambda_{c}^{(s)}
\end{equation}
and by defining:
\begin{equation} 
\delta_{EW}=\frac{-\frac{3}{2}\left(\frac{c_{9}+c_{10}}{c_{1}+c_{2}}\right)}
{1+\frac{3}{2}\left(\frac{c_{9}+c_{10}}{c_{1}+c_{2}}\right)}\cdot\frac{\lambda^{(s)}_{c}}{|\lambda^{(s)}_{u}|}\approx 0.65
\end{equation}
($\lambda_{c}^{(s)}$ is real up to third order in $|V_{us}|$), 
they obtained :
($1+\frac{3}{2}\left(\frac{c_{9}+c_{10}}{c_{1}+c_{2}}\right)\approx 1.01$)
\begin{eqnarray}
\label{phases} 
\sqrt{2}{\cal A}(B^{+} \rightarrow K^{+} \pi^{0})+
{\cal A}(B^{+} \rightarrow K^{0} \pi^{+})
&=&-|\lambda_{u}^{(s)}|(e^{i\gamma}-\delta_{EW})
\left[1+\frac{3}{2}\left(\frac{c_{9}+c_{10}}{c_{1}+c_{2}}\right)\right]
(T+C) \nonumber \\
&\approx&-|\lambda_{u}^{(s)}|(e^{i\gamma}-\delta_{EW})(T+C) 
\end{eqnarray}
Let us look at the CP averaged ratio:
\begin{equation}
R=\frac{2[Br(B^{+} \rightarrow K^{+} \pi^{0})+
Br(B^{-} \rightarrow K^{-} \pi^{0})]}
{Br(B^{+} \rightarrow K^{0} \pi^{+})+
Br(B^{-} \rightarrow \overline{K^{0}} \pi^{-})},
\end{equation}
which can be written as:
\begin{equation}
R=\frac{|\lambda_{u}^{(s)}(P_{uc}+A)+\lambda_{t}^{(s)}(P_{ct}+P^{EW})
-\lambda_{u}^{(s)}(e^{i\gamma}-\delta_{EW})(T+C)|^2+CP}
{|\lambda_{u}^{(s)}(P_{uc}+A)+\lambda_{t}^{(s)}(P_{ct}+P^{EW})|^2+CP}
\end{equation}
where $P^{EW}=P^{EW}(B^{+} \rightarrow K^{0} \pi^{+})$ and ``$CP$ '' denotes the CP conjugate amplitude. By using the orthogonality relation (\ref{ortho}) we can write this ratio as:
\begin{equation}
R=\frac{|\lambda_{u}^{(s)}[(P_{uc}+A)-(P_{ct}+P^{EW})]-\lambda_{c}^{(s)}(P_{ct}+P^{EW})-\lambda_{u}^{(s)}(e^{i\gamma}-\delta_{EW})(T+C)|^2+CP}
{|\lambda_{u}^{(s)}[(P_{uc}+A)-(P_{ct}+P^{EW})]-\lambda_{c}^{(s)}(P_{ct}+P^{EW})|^2+CP}.
\end{equation}
If we define:
\begin{equation}
\epsilon e^{i\phi_{T}}=\left|\frac{\lambda_{u}^{(s)}}{\lambda_{c}^{(s)}}\right|
\frac{T+C}{\left|P_{ct}+P^{EW}\right|}
\end{equation}
\begin{equation}
e^{i\phi_{P}}=\left|\frac{\lambda_{u}^{(s)}}{\lambda_{c}^{(s)}}\right|
\frac{P_{ct}+P^{EW}}{\left|P_{ct}+P^{EW}\right|}
\end{equation}
\begin{equation}
\epsilon_{A} e^{i\phi_{A}}=\left|\frac{\lambda_{u}^{(s)}}{\lambda_{c}^{(s)}}\right|
\left[-\frac{P_{uc}+A}{\left|P_{ct}+P^{EW}\right|}+
\frac{P_{ct}+P^{EW}}{\left|P_{ct}+P^{EW}\right|}\right],
\end{equation}
we find:
\begin{equation}
\label{rate}
R=\frac{\left|\epsilon e^{i\phi_{T}}(e^{i\gamma}-\delta_{EW})
-\epsilon_{A} e^{i\phi_{A}}e^{i\gamma}-e^{i\phi_{P}}\right|^{2}
+\left|\epsilon e^{i\phi_{T}}(e^{-i\gamma}-\delta_{EW})
-\epsilon_{A} e^{i\phi_{A}}e^{-i\gamma}-e^{i\phi_{P}}\right|^{2}}
{\left|\epsilon_{A} e^{i\phi_{A}}e^{i\gamma}+e^{i\phi_{P}}\right|^{2}+
\left|\epsilon_{A} e^{i\phi_{A}}e^{-i\gamma}+e^{i\phi_{P}}\right|^{2}}.
\end{equation}
We are interested only in the lowest order in $\epsilon$, therefore expanding  $R$ we find:
\begin{eqnarray}
R&=&1-\frac{ \epsilon e^{i\Delta\phi}(e^{i\gamma}-\delta_{EW})
+\epsilon e^{-i\Delta\phi}(e^{-i\gamma}-\delta_{EW})
+\epsilon e^{i\Delta\phi}(e^{-i\gamma}-\delta_{EW})
+\epsilon e^{-i\Delta\phi}(e^{i\gamma}-\delta_{EW})}
{\left|\epsilon_{A} e^{i\phi_{A}}e^{i\gamma}+e^{i\phi_{P}}\right|^{2}+
\left|\epsilon_{A} e^{i\phi_{A}}e^{-i\gamma}+e^{i\phi_{P}}\right|^{2}}.
\nonumber \\
&& \nonumber \\
&=&1-2\epsilon \cos \Delta\phi(\cos \gamma-\delta_{EW})+
{\cal O}(\epsilon^{2})+{\cal O}(\epsilon\epsilon_{A})
+{\cal O}(\epsilon_{A}^{2}),\;\; \Delta\phi=\phi_{T}-\phi_{A}.
\end{eqnarray}
Which leads to the constraint:
\begin{equation}
|\cos \gamma - \delta_{EW}|\geq
\frac{|1-R|}{2\epsilon}.
\end{equation}
In order to relate the expansion parameter:
\begin{equation}
\label{small}
\epsilon=\left|\frac{\lambda_{u}^{(s)}}{\lambda_{c}^{(s)}}\right|
\frac{\left|T+C\right|}{\left|P_{ct}+P^{EW}\right|}
\end{equation}
to measurable quantities we note that:
\begin{eqnarray}
\sqrt{2}\left|{\cal A}(B^{+} \rightarrow \pi^{0} \pi^{+})\right| &= &
|\lambda_{u}^{(d)}| |T+C|, \nonumber \\
\left|{\cal A}(B^{+} \rightarrow K^{0} \pi^{+})\right| & \approx &
|\lambda_{c}^{(s)}|P_{ct}+P^{EW}|, 
\end{eqnarray}
so that:
\begin{eqnarray}
\epsilon&=&\sqrt{2}\left|\frac{{\cal A}(B^{+}\rightarrow\pi^{0}\pi^{+})}
{{\cal A}(B^{+} \rightarrow K^{0} \pi^{+})}\right|
\left|\frac{\lambda_{u}^{(s)}}{\lambda_{u}^{(d)}}\right| \nonumber \\
&=&\sqrt{2}\left|\frac{{\cal A}(B^{+}\rightarrow\pi^{0}\pi^{+})}
{{\cal A}(B^{+} \rightarrow K^{0} \pi^{+})}\right|
\left|\frac{V_{us}}{V_{ud}}\right|.
\end{eqnarray}
$SU(3)$ breaking affects appear only when we relate the ``$T+C$'' in 
${\cal A}(B^{+}\rightarrow\pi^{0}\pi^{+})$ to``$T+C$'' in 
${\cal A}(B^{+} \rightarrow K^{+} \pi^{0})$. Using the factorization approximation it can be shown that the two are related by the ratio of the decay constants $f_{\pi}$ and $f_{K}$ \cite{nr,grl}. Including this ratio we have:
\begin{equation}
\epsilon=\sqrt{2}\,\frac{f_{K}}{f_{\pi}}\left|\frac{{\cal A}(B^{+}\rightarrow\pi^{0}\pi^{+})}{{\cal A}(B^{+} \rightarrow K^{0} \pi^{+})}\right|
\left|\frac{V_{us}}{V_{ud}}\right|.
\end{equation}
 From the current data the value of $\epsilon$ is \cite{pdg}: $0.09\pm 0.05$ \cite{babar1} (BaBar), $0.15\pm 0.07$ \cite{belle1} (Belle), $0.13\pm 0.11$ \cite{cassel} (CLEO). \\

There seem to be four essential features to the Neubert Rosner method:
\begin{enumerate}
\item The EWP contribution ($Q_{9}$,$Q_{10}$) can be related to the Tree contribution (see equation \ref{ewpt})
\item There are two amplitudes that differ in a complex number (see equation \ref{phases}), which has one strong phase (the phase of $T+C$) and a unique weak phase ($e^{i\gamma}-\delta_{EW}$).
\item This number can be measured through some decays (see equation \ref{measure}) .
\item The difference between the amplitudes is small compared to (one of) the amplitudes (see equation \ref{small}). 
\end{enumerate}
As we shall see in the following section the first requirement  would be easy to fulfill and in fact we will find that the EWP contribution can be related in a
systematic way to the Tree contribution. The third requirement will turn out to be  the hardest to fulfill.\\
\subsection{Decomposition of the effective Hamiltonian}
The effective Hamiltonian for charmless $B$ decays is composed of two main parts $\Delta S=1$ and $ \Delta S=0$. Each part is composed from the 
operators $ Q_{1}$,\ldots, $ Q_{10}$. We will find it useful to ``rearrange'' the various parts of the Hamiltonian before embarking on the $SU(3)$ decomposition.\\
Lets look first at the  $ \Delta S=1$ part, it can be written as:\\
\begin{equation}
{\cal H}_{\rm eff}=\frac{G_{F}}{\sqrt{2}}\left[\lambda^{(s)}_{u}
(c_{1}Q^{(u)}_{1}+c_{2}Q^{(u)}_{2})+
\lambda^{(s)}_{c}(c_{1}Q^{(c)}_{1}+c_{2}Q^{(c)}_{2})+
(\lambda^{(s)}_{u}+\lambda^{(s)}_{c})
\sum_{i=3}^{10}c_{i}Q_{i}\right]
\end{equation} 
Where $ Q_{1}$ and $ Q_{2}$ are the ``Tree'' operators:
\begin{eqnarray}
Q_{1}^{(u)} & = & (\overline{b}_{i}u_{j})_{V-A}
(\overline{u}_{j}s_{i})_{V-A}  =  (\overline{b}_{i}s_{i})_{V-A}
(\overline{u}_{j}u_{j})_{V-A} \nonumber \\
Q_{1}^{(c)} & = & (\overline{b}_{i}c_{j})_{V-A}
(\overline{c}_{j}s_{i})_{V-A}  =  (\overline{b}_{i}s_{i})_{V-A}
(\overline{c}_{j}c_{j})_{V-A} \nonumber \\
Q_{2}^{(u)}&=&(\overline{b}_{i}u_{i})_{V-A}
(\overline{u}_{j}s_{j})_{V-A}  \nonumber  \\
Q_{2}^{(c)}&=&(\overline{b}_{i}u_{i})_{V-A}
(\overline{u}_{j}s_{j})_{V-A} \nonumber . \\
\end{eqnarray}
$ Q_{3}$, \ldots, $ Q_{6}$ are the ``Penguin'' operators:
\begin{eqnarray}
Q_{3} & = & (\overline{b}_{i}s_{i})_{V-A}
\sum_{q=u,d,s,c}(\overline{q}_{j}q_{j})_{V-A}   \nonumber \\
Q_{4} & =& (\overline{b}_{i}s_{j})_{V-A} 
\sum_{q=u,d,s,c}(\overline{q}_{j}q_{i})_{V-A}
 =  (\overline{b}_{i}q_{i})_{V-A} 
\sum_{q=u,d,s,c}(\overline{s}_{j}q_{j})_{V-A} \nonumber \\
Q_{5} & = & (\overline{b}_{i}s_{i})_{V-A}
\sum_{q=u,d,s,c}(\overline{q}_{j}q_{j})_{V+A}   \nonumber \\
Q_{6} & =& (\overline{b}_{i}s_{j})_{V-A} 
\sum_{q=u,d,s,c}(\overline{q}_{j}q_{i})_{V-A}
 =  (\overline{b}_{i}q_{i})_{V-A} 
\sum_{q=u,d,s,c}(\overline{s}_{j}q_{j})_{V+A} \nonumber. \\
\end{eqnarray}
Out of the four ``EWP'' (i.e.``ElectroWeak Penguins'') Operators: 
$ Q_{7}$, \ldots, $ Q_{10}$, we can neglect $ Q_{7}$ and $ Q_{8}$:
\begin{eqnarray}
Q_{7} & = &\frac{3}{2} (\overline{b}_{i}s_{i})_{V-A}
\sum_{q=u,d,s,c}e_{q}(\overline{q}_{j}q_{j})_{V+A}   \nonumber \\
Q_{8} & =&\frac{3}{2} (\overline{b}_{i}s_{j})_{V-A} 
\sum_{q=u,d,s,c}e_{q}(\overline{q}_{j}q_{i})_{V+A} \\
\end{eqnarray}
because of the smallness of $ c_{7}$ and  $ c_{8}$ with respect to the 
other Wilson Coefficients.\\ 
Thus\cite{buras}: $|c_{7}/c_{9}|\leq 0.04$, $|c_{7}/c_{1}|\leq 0.002$, 
$|c_{8}/c_{10}|\leq 0.3$ and $|c_{8}/c_{1}|\leq 0.003$. \\
The remaining ``EWP'' operators are:
\begin{eqnarray}
Q_{9} & = &\frac{3}{2} (\overline{b}_{i}s_{i})_{V-A}
\sum_{q=u,d,s,c}e_{q}(\overline{q}_{j}q_{j})_{V-A}   \nonumber \\
Q_{10} & =&\frac{3}{2} (\overline{b}_{i}s_{j})_{V-A} 
\sum_{q=u,d,s,c}e_{q}(\overline{q}_{j}q_{i})_{V-A}
 =  (\overline{b}_{i}q_{i})_{V-A} 
\sum_{q=u,d,s,c}e_{q}(\overline{s}_{j}q_{j})_{V-A} \nonumber \\
\end{eqnarray}
For some of the operators we have used the Fiertz transformations for {\it
anti-commuting fermion fields}, which are \cite{fiertz}: 
\begin{equation}
(\overline{\alpha}\beta)_{V\pm A}(\overline{\gamma}\delta)_{V \pm A}=
(\overline{\gamma}\beta)_{V\pm A}(\overline{\alpha}\delta)_{V \pm A} 
\end{equation}
The Operator $Q_{9}$ can be rearranged in the following way (we suppress the color indices and the chirality structure for the moment):
\begin{eqnarray}
Q_{9}&=&\frac{3}{2}(\overline{b}s)
\sum_{q=u,d,s,c}e_{q}(\overline{q}q)\nonumber \\
 &=&\frac{3}{2}\overline{b}s
\left(\frac{2}{3}\overline{u}u
-\frac{1}{3}\overline{d}d
-\frac{1}{3}\overline{s}s
+\frac{2}{3}\overline{c}c\right)\nonumber \\
&=&\frac{3}{2}\overline{b}s
\left(-\frac{1}{3}\overline{u}u
-\frac{1}{3}\overline{d}d
-\frac{1}{3}\overline{s}s
+\overline{u}u+\frac{2}{3}\overline{c}c\right)\nonumber \\
 &=& \left[-\frac{1}{2}\overline{b}s(\overline{u}u+\overline{d}d
+\overline{s}s)\right]+\left[\frac{3}{2}\overline{b}s\overline{u}u
+\overline{b}s\overline{c}c\right]
\end{eqnarray}
In this decomposition $Q_{9}$ is made from two parts: one resembles
$Q_{3}$ and the other resembles $Q_{1}$. Therefore it seems useful to attach these parts to $Q_{3}$ and $Q_{1}$ respectively before the $SU(3)$ decomposition. A similar operation can be applied to $Q_{10}$. Thus, after the Fiertz transformation we can write the effective Hamiltonian as (we suppress the color indices):
\begin{eqnarray}
\frac{\sqrt{2}{\cal H}_{\rm eff}}{G_{F}}& = &
(\overline{b}s\overline{u}u)_{V-A,V-A}[\lambda^{(s)}_{u}c_{1}+
(\lambda^{(s)}_{u}+\lambda^{(s)}_{c})\frac{3}{2}c_{9}]
\nonumber \\
& + & (\overline{b}u\overline{u}s)_{V-A,V-A}[\lambda^{(s)}_{u}c_{2}+
(\lambda^{(s)}_{u}+\lambda^{(s)}_{c})\frac{3}{2}c_{10}]
\nonumber \\
& + & (\overline{b}s\overline{c}c)_{V-A,V-A}[\lambda^{(s)}_{c}c_{1}+
(\lambda^{(s)}_{u}+\lambda^{(s)}_{c})(c_{3}+c_{9})]
\nonumber \\
& + & (\overline{b}c\overline{c}s)_{V-A,V-A}[\lambda^{(s)}_{c}c_{2}+
(\lambda^{(s)}_{u}+\lambda^{(s)}_{c})( c_{4}+c_{10} )
] \nonumber \\
& + & (\overline{b}s\overline{u}u+\overline{b}s\overline{d}d+
\overline{b}s\overline{s}s)_{V-A,V-A}
[(\lambda^{(s)}_{u}+\lambda^{(s)}_{c})( c_{3}-
\frac{1}{2}c_{9})]\nonumber \\
& + &(\overline{b}u\overline{u}s+
\overline{b}d\overline{d}s+
\overline{b}s\overline{s}s)_{V-A,V-A}[(\lambda^{(s)}_{u}+
\lambda^{(s)}_{c})(c_{4}-\frac{1}{2}c_{10})]
\nonumber \\
& + & (\overline{b}s\overline{c}c)_{V-A,V+A}
(\lambda^{(s)}_{u}+\lambda^{(s)}_{c})c_{5}
\nonumber \\ 
& +& (\overline{b}c\overline{c}s)_{V-A,V+A}
(\lambda^{(s)}_{u}+\lambda^{(s)}_{c})c_{6}
 \nonumber \\
& + & (\overline{b}s\overline{u}u+\overline{b}s\overline{d}d+
\overline{b}s\overline{s}s)_{V-A,V+A}
(\lambda^{(s)}_{u}+\lambda^{(s)}_{c})c_{5}
\nonumber \\
& +&(\overline{b}u\overline{u}s+
\overline{b}d\overline{d}s+
\overline{b}s\overline{s}s)_{V-A,V+A}(\lambda^{(s)}_{u}+
\lambda^{(s)}_{c})c_{6}
\nonumber \\
\end{eqnarray}

\newpage
\noindent ${\cal H}_{\rm eff}$  is a  linear combinations of four quark operators of the form $(\overline{b}q_{1})(\overline{q_{2}}q_{3})$.
These operators transform as $\overline{3}\bigotimes 3 \bigotimes \overline{3}$
under $SU(3)$-flavor and can be decomposed into sums of irreducible operators :
$\overline{15}\,$,$\,6\,$,$\,\overline{3^{(a)}}\,$,$\,\overline{3^{(s)}}\,$ where  the index: 'a' ('s') designates antisymmetry (symmetry) under the
 interchange of $\overline{q_{1}}$ and $\overline{q_{3}}$. The tensor product: $\overline{3}\bigotimes 3 \bigotimes \overline{3}$ is unitary equivalent to $\overline{3}\bigotimes\overline{3}\bigotimes 3$ \cite{operator}. Therefore the $SU(3)$ decomposition can be done in either form without changing the physics. In the form $\overline{3}\bigotimes\overline{3}\bigotimes 3$ the ``interchange symmetry'' is manifest, therefore the four quark operators will be written as $\overline{q_{1}}\,\overline{q_{3}}q_{2}$.

\noindent In the following table the four quark operators, which appear in ${\cal H}_{\rm eff}$,  are decomposed (using $SU(3)$ Clebsch-Gordan tables \cite{kaeding}). {\it Notice that in the Hamiltonian the operators appear as $q_{1}\,\overline{q_{2}}\,q_{3}$ and in the table they appear as $\overline{q_{1}}\,\overline{q_{3}}q_{2}$}. 

\begin{table}[h]
\begin{tabular}{|c|c|c|c|c|c|c|c|c|c|c|}	\hline

\,&$\overline{15}_{I=1}$&$\overline{15}_{I=0}$&$6_{I=1}$&$\overline{3}^{(a)}_{I=0}$&$\overline{3}^{(s)}_{I=0}$&$\overline{15}_{I=3/2}$&$\overline{15}_{I=1/2}$&$6_{I=1/2}$&$\overline{3}^{(a)}_{I=1/2}$&$\overline{3}^{(s)}_{I=1/2}$ \\ \hline
$\overline{d}\,\overline{d}\,d$&&&&&&$\sqrt{1/3}$ &$-\sqrt{1/6}$&&&$\sqrt{1/2}$ \\ \hline
$\overline{d}\,\overline{u}\,u$&&&&&&$-\sqrt{1/3}$ &$-\sqrt{1/24}$&$-1/2$&$1/2$&$\sqrt{1/8}$ \\ \hline
$\overline{d}\,\overline{s}\,d$&$1/2$&$-\sqrt{1/8}$&$1/2$&$-1/2$&$\sqrt{1/8}$& &&&& \\ \hline
$\overline{d}\,\overline{s}\,s$&&&&&&&$\sqrt{3/8}$&$1/2$&$1/2$&$\sqrt{1/8}$ \\ \hline
$\overline{u}\,\overline{d}\,u$&&&&&&$-\sqrt{1/3}$ &$-\sqrt{1/24}$&$1/2$&$-1/2$&$\sqrt{1/8}$ \\ \hline
$\overline{u}\,\overline{s}\,u$&$-1/2$&$-\sqrt{1/8}$&$-1/2$&$-1/2$&$\sqrt{1/8}$& &&&& \\ \hline
$\overline{s}\,\overline{d}\,d$&$1/2$&$-\sqrt{1/8}$&$-1/2$&$1/2$&$\sqrt{1/8}$& &&&& \\ \hline
$\overline{s}\,\overline{d}\,s$&&&&&&&$\sqrt{3/8}$&$-1/2$&$-1/2$&$\sqrt{1/8}$ \\ \hline
$\overline{s}\,\overline{u}\,u$&$-1/2$&$-\sqrt{1/8}$&$1/2$&$1/2$&$\sqrt{1/8}$& &&&& \\ \hline
$\overline{s}\,\overline{s}\,s$&&$\sqrt{1/2}$&&&$\sqrt{1/2}$& &&&& \\ \hline

\end{tabular}
\caption{Operator Decomposition}
\end{table}

In the table the hypercharge $ Y$ and the $ I_{3}$ indices of the
operators were suppressed.\\ 
If $I$ is an integer the indices are $ Y=2/3,I_{3}=0$.\\
If $I$ is an half-integer the indices are $ Y=-1/3,I_{3}=1/2$.\\ 

The various ``charming penguins'' operators, that is, operators of the form 
$\overline{b}s\overline{c}c$ and $\overline{b}c\overline{c}s$, do not appear in the table above. But they are all $\overline{3}_{I=0}$ operators since the operator $s$ transforms as $\overline{3}$. The $SU(3)$ quantum numbers of this operator are $Y=2/3\,\, I=0\,\, I_{3}=0$.  
\newpage
Using the decomposition we can write the effective Hamiltonian as:
\begin{eqnarray}
\frac{\sqrt{2}{\cal H}_{\rm eff}}{G_{F}}& = &
-\frac{1}{2}(\overline{15}_{I=1}+\frac{1}{\sqrt{2}}\overline{15}_{I=0})
[\lambda^{(s)}_{u}(c_{2}+c_{1})+\frac{3}{2}(\lambda^{(s)}_{u}+\lambda^{(s)}_{c})(c_{9}+c_{10})]\nonumber \\
& - & \frac{1}{2}\cdot6_{I=1}[\lambda^{(s)}_{u}(c_{2}-c_{1})-\frac{3}{2}(\lambda^{(s)}_{u}+\lambda^{(s)}_{c})(c_{9}-c_{10})]\nonumber \\
& + & \frac{1}{\sqrt{8}}\cdot\overline{3}^{(s)}_{I=0}
[\lambda^{(s)}_{u}(c_{2}+c_{1})+\frac{3}{2}(\lambda^{(s)}_{u}+\lambda^{(s)}_{c})(c_{9}+c_{10})]\nonumber \\
& - & \frac{1}{2}\cdot\overline{3}^{(a)}_{I=0}[\lambda^{(s)}_{u}(c_{2}-c_{1})-\frac{3}{2}(\lambda^{(s)}_{u}+\lambda^{(s)}_{c})(c_{9}-c_{10})]\nonumber \\
& + & \overline{3}_{I=0}[\lambda^{(s)}_{c}(c_{2}+c_{1})+
(\lambda^{(s)}_{u}+\lambda^{(s)}_{c})(c_{3}+c_{9}+c_{4}+c_{10})] \nonumber \\
& + & \sqrt{2}\cdot \overline{3}^{(s)}_{I=0}(\lambda^{(s)}_{u}+\lambda^{(s)}_{c})[c_{3}+c_{4}-\frac{1}{2}(c_{9}+c_{10})]\nonumber \\
& + & \overline{3}^{(a)}_{I=0}(\lambda^{(s)}_{u}+\lambda^{(s)}_{c})[c_{3}-c_{4}-\frac{1}{2}(c_{9}-c_{10})]\nonumber \\
& + & (\lambda^{(s)}_{u}+\lambda^{(s)}_{c})[\overline{3}_{I=0}(c_{5}+c_{6})
+\sqrt{2}\cdot \overline{3}^{(s)}_{I=0}(c_{5}+c_{6})+\overline{3}^{(a)}_{I=0}
(c_{5}-c_{6})] \nonumber \\
\end{eqnarray}
As can be seen from the decomposition, the main contribution to the effective Hamiltonian are the several ``$\overline{3}$'' operators.
A priori, that is before the factorization approximation, the dominant 
operators would be the ones with the largest coefficients. The hierarchy
of the coefficients is:
\begin{displaymath}
|\lambda^{(s)}_{u}(c_{2}\pm c_{1})|,|\lambda^{(s)}_{c}(c_{9}\pm c_{10})|\ll
|\lambda^{(s)}_{c}(c_{3}\pm c_{4})|,|\lambda^{(s)}_{c}(c_{5}\pm c_{6})|\ll
|\lambda^{(s)}_{c}(c_{2}\pm c_{1})|
\end{displaymath}
Where $\ll$ denotes an order of magnitude.\\
Therefore, A priori, the matrix elements of the ``$6$'' and ``$\overline{15}$''operators would be two orders of magnitude  smaller then the largest matrix element of the ``$\overline{3}$'' operators.
Since all of the ``$\overline{3}$'' operators, would contribute to the amplitude similar reduced matrix element, namely, $<8\|\overline{3}\|3>$ and $<1\|\overline{3}\|3>$, we can write the effective Hamiltonian as:
\begin{eqnarray}
\frac{\sqrt{2}{\cal H}_{\rm eff}}{G_{F}}& = &
-\frac{1}{2}(\overline{15}_{I=1}+\frac{1}{\sqrt{2}}\overline{15}_{I=0})
\lambda^{(s)}_{u}(c_{2}+c_{1})[
1+\frac{3}{2}\cdot\frac{c_{9}+c_{10}}{c_{2}+c_{1}}+\frac{3}{2}\cdot\frac{\lambda^{(s)}_{c}}{\lambda^{(s)}_{u}}\cdot\frac{c_{9}+c_{10}}{c_{2}+c_{1}}]
\nonumber \\
& - & \frac{1}{2}\cdot6_{I=1}\lambda^{(s)}_{u}(c_{2}-c_{1})[
1-\frac{3}{2}\cdot\frac{c_{9}-c_{10}}{c_{2}-c_{1}}-\frac{3}{2}\cdot\frac{\lambda^{(s)}_{c}}{\lambda^{(s)}_{u}}\cdot\frac{c_{9}-c_{10}}{c_{2}-c_{1}}]\nonumber \\
& + &\lambda^{(s)}_{u}a_{u}\cdot\overline{3}_{I=0}+\lambda^{(s)}_{c}a_{c}\cdot\overline{3}_{I=0}\nonumber, \\
\end{eqnarray}
where $a_{u}$ and $a_{c}$ are the appropriate linear combinations of the Wilson coefficients (More explicit expressions appear in Appendix B).\\ 
At a scale of $\mu=m_{b}$ \cite{buras}
\begin{equation}
c_{1}=-0.308\;\;c_{2}=1.144\;\;c_{9}=-1.280\alpha\;\;c_{10}=0.328\alpha,
\end{equation}
where $\alpha=1/129$, and therefore:
\begin{equation}
\frac{c_{9}+c_{10}}{c_{2}+c_{1}}=-1.139\alpha,\;\;
\frac{c_{9}-c_{10}}{c_{2}-c_{1}}=-1.107\alpha.
\end{equation}
Let $\kappa$ denote the average of the two ratios:
\begin{equation} 
\kappa\approx\frac{c_{9}-c_{10}}{c_{2}-c_{1}}
\approx\frac{c_{9}+c_{10}}{c_{2}+c_{1}}\approx-1.12\alpha.
\end{equation}
Using $\kappa$ and $\lambda^{(s)}_{u}=|\lambda^{(s)}_{u}|e^{i\gamma}$ we can write the effective Hamiltonian as:
\begin{eqnarray}
\frac{\sqrt{2}{\cal H}_{\rm eff}}{G_{F}}& = &
-\frac{1}{2}(\overline{15}_{I=1}+\frac{1}{\sqrt{2}}\overline{15}_{I=0})
|\lambda^{(s)}_{u}|(c_{2}+c_{1})(1+\frac{3}{2}\kappa)[e^{i\gamma}+
\frac{\frac{3}{2}\kappa}{1+\frac{3}{2}\kappa}\cdot
\frac{\lambda^{(s)}_{c}}{|\lambda^{(s)}_{u}|}]\nonumber \\
& - & \frac{1}{2}\cdot6_{I=1}|\lambda^{(s)}_{u}|(c_{2}-c_{1})(1-\frac{3}{2}
\kappa)[e^{i\gamma}-\frac{\frac{3}{2}\kappa}{1-\frac{3}{2}\kappa}\cdot
\frac{\lambda^{(s)}_{c}}{|\lambda^{(s)}_{u}|}]\nonumber \\
& + &\lambda^{(s)}_{u}a_{u}\cdot\overline{3}_{I=0}+\lambda^{(s)}_{c}a_{c}\cdot\overline{3}_{I=0}\nonumber \\
\end{eqnarray} 
Since $1+\frac{3}{2}\kappa=1.01$ and $1-\frac{3}{2}\kappa=0.99$ we can 
approximate $1\pm\frac{3}{2}\kappa\approx1$ and define:
\begin{equation} 
\delta_{EW}=-\frac{\frac{3}{2}\kappa}{1\pm\frac{3}{2}\kappa}\cdot
\frac{\lambda^{(s)}_{c}}{|\lambda^{(s)}_{u}|}
\approx 0.65
\end{equation}
This enables us to write the effective Hamiltonian in a very compact form:
\begin{eqnarray}
\label{hcomp1}
\frac{\sqrt{2}{\cal H}_{\rm eff}}{G_{F}}(\Delta S=1)& = &
-\frac{1}{2}(\overline{15}_{I=1}+\frac{1}{\sqrt{2}}\overline{15}_{I=0})
|\lambda^{(s)}_{u}|(c_{2}+c_{1})(e^{i\gamma}-\delta_{EW})
\nonumber \\
& - & \frac{1}{2}\cdot6_{I=1}|\lambda^{(s)}_{u}|(c_{2}-c_{1})
(e^{i\gamma}+\delta_{EW}) \nonumber \\
& + &\lambda^{(s)}_{u}a_{u}\cdot\overline{3}_{I=0}+\lambda^{(s)}_{c}a_{c}
\cdot\overline{3}_{I=0}\nonumber.\\
\end{eqnarray} 
We can write the $\Delta S= 0 $ part of the effective Hamiltonian in a similar way:
\begin{eqnarray}
\label{hcomp0}
\frac{\sqrt{2}{\cal H}_{\rm eff}}{G_{F}}(\Delta S=0)& = &
-\frac{1}{2}(\frac{2}{\sqrt{3}}\overline{15}_{I=3/2}+\frac{1}{\sqrt{6}}\overline{15}_{I=1/2})
|\lambda^{(d)}_{u}|(c_{2}+c_{1})(e^{i\gamma}-\bar{\delta}_{EW})
\nonumber \\
& - & \frac{1}{2}\cdot(-6_{I=1/2})|\lambda^{(d)}_{u}|(c_{2}-c_{1})
(e^{i\gamma}+\bar{\delta}_{EW}) \nonumber \\
& + &\lambda^{(d)}_{u}a_{u}\cdot\overline{3}_{I=1/2}+\lambda^{(d)}_{c}a_{c}
\cdot\overline{3}_{I=1/2}\nonumber, \\
\end{eqnarray}
where we have defined:
\begin{equation} 
\bar{\delta}_{EW}=-\frac{\frac{3}{2}\kappa}{1\pm\frac{3}{2}\kappa}\cdot
\frac{\lambda^{(d)}_{c}}{|\lambda^{(d)}_{u}|}
\approx 0.03.
\end{equation}
The fact that $\bar{\delta}_{EW}$ is so small compared to $\delta_{EW}$ is just another way of saying that the contribution of the electroweak penguins is negligible in $\Delta S= 0 $ decays. We see that the EWP effects appear in the Hamiltonian in the ``$\delta_{EW}$ contribution''  and in modifying  $a_{u}$ and $a_{c}$.\\
It should be noted that the coefficients $a_{u}$, $a_{c}$ are the same for $\Delta S= 0 $ and the $\Delta S= 1$ parts of the Hamiltonian.
{\it In the following we will suppress the factor $\frac{G_{F}}{\sqrt{2}}$, which appear in all of the amplitudes.}  
Having rearranged the Effective Hamiltonian we are ready to decompose the various decay amplitudes according to Wigner-Eckart theorem \cite{jj} using $SU(3)$ Clebsch-Gordan tables \cite{kaeding}. The results of the decomposition appear in appendix A.

\section{Generalization of the Neubert-Rosner Method}
 
\subsection{The Neubert-Rosner Method}
Now, we are going to repeat the Neubert-Rosner calculation using our analysis. From the tables in appendix A we find:
\begin{equation} 
\label{nram}
\sqrt{2}{\cal A}(B^{+} \rightarrow K^{+} \pi^{0})+
{\cal A}(B^{+} \rightarrow K^{0} \pi^{+})=20 a'_{5},
\end{equation}
while:
\begin{equation} 
\sqrt{2}{\cal A}(B^{+} \rightarrow \pi^{0} \pi^{+})=20 a_{5}.
\end{equation}
In order to use this relation in the expansion of the CP-averaged rate $R$,
one has to identify the ``large'' parts of the amplitudes. Generally, we can 
write:
\begin{eqnarray}
\sqrt{2}{\cal A}(B^{+} \rightarrow \pi^{0} \pi^{+})
&=& +|\lambda_{u}^{(d)}|e^{i\gamma}B \nonumber \\
\sqrt{2}{\cal A}(B^{+} \rightarrow K^{+} \pi^{0})
&=& +\lambda_{u}^{(s)}A_{u}+\lambda_{c}^{(s)}A_{c}+
|\lambda_{u}^{(s)}|(e^{i\gamma}-\delta_{EW})B \\
{\cal A}(B^{+} \rightarrow K^{0} \pi^{+})
&=& -\lambda_{u}^{(s)}A_{u}-\lambda_{c}^{(s)}A_{c}\nonumber 
\end{eqnarray}
where we have defined:
\pagebreak
\begin{eqnarray}
B&=&-20\cdot\frac{1}{2}(c_{2}+c_{1})
\sqrt{\frac{1}{200}}\langle{27}\|\overline{15}\|3\rangle\nonumber \\  
A_{u}&=&+3\cdot a_{u}\sqrt{\frac{1}{15}}\langle{8}_{S}\|\overline{3}\|3\rangle) \nonumber \\
&&-\frac{1}{2}(c_{2}-c_{1})\sqrt{\frac{1}{5}}\langle{8}_{S}\|6\|3\rangle
\nonumber\\
&&+3\cdot\frac{1}{2}(c_{2}+c_{1})\sqrt{\frac{1}{50}}\langle{8}_{S}\|\overline{15}\|3\rangle\nonumber\\
&&+4\cdot\frac{1}{2}(c_{2}+c_{1})\sqrt{\frac{1}{200}}\langle{27}\|\overline{15}\|3\rangle\nonumber\\
A_{c}&=&
+3\cdot a_{c}\sqrt{\frac{1}{15}}\langle{8}_{S}\|\overline{3}\|3\rangle) \nonumber \\
&&-\frac{1}{2}\frac{|\lambda^{(s)}_{u}|\delta_{EW}}{\lambda^{(s)}_{c}}(c_{2}-c_{1})\sqrt{\frac{1}{5}}\langle{8}_{S}\|6\|3\rangle
\nonumber\\
&&+3\cdot\frac{1}{2}\frac{|\lambda^{(s)}_{u}|\delta_{EW}}{\lambda^{(s)}_{c}}(c_{2}+c_{1})\sqrt{\frac{1}{50}}\langle{8}_{S}\|\overline{15}\|3\rangle\nonumber\\
&&+4\cdot\frac{1}{2}\frac{|\lambda^{(s)}_{u}|\delta_{EW}}{\lambda^{(s)}_{c}}(c_{2}+c_{1})\sqrt{\frac{1}{200}}\langle{27}\|\overline{15}\|3\rangle.
\end{eqnarray}
Since \cite{pdg}
\begin{equation}
0.01\leq\left|\frac{\lambda^{(s)}_{u}}{\lambda^{(s)}_{c}}\right|\leq0.03\, ,
\end{equation}
we can see that the term $\lambda_{c}^{(s)}A_{c}$ is the ``large'' part of the amplitude. This is equivalent to the statement that the dominant contribution in $\Delta S=1$ decays is the ``Penguin'' contribution \cite{nr}.\\
Therefore our expansion parameter would be :
\begin{equation}
\epsilon=\left|\frac{\lambda^{(s)}_{u}}{\lambda^{(s)}_{c}}\right|
\left|\frac{B}{A_{c}}\right|.
\end{equation}
We now return to the CP averaged ratio:
\begin{eqnarray}
R&=&\frac{2[Br(B^{+} \rightarrow K^{+} \pi^{0})+
Br(B^{-} \rightarrow K^{-} \pi^{0})]}
{Br(B^{+} \rightarrow K^{0} \pi^{+})+
Br(B^{-} \rightarrow \overline{K^{0}} \pi^{-})} \nonumber \\
&&\nonumber \\
&=&\frac{|\lambda_{u}^{(s)}A_{u}+\lambda_{c}^{(s)}A_{c}
+|\lambda_{u}^{(s)}|(e^{i\gamma}-\delta_{EW})B|^{2}+CP}
{|\lambda_{u}^{(s)}A_{u}+\lambda_{c}^{(s)}A_{c}|^{2}+CP},
\end{eqnarray}
If we define:
\begin{equation}
\epsilon e^{i\phi_{T}}=\left|\frac{\lambda_{u}^{(s)}}{\lambda_{c}^{(s)}}\right|
\frac{B}{A_{c}}
\end{equation}
\begin{equation}
e^{i\phi_{P}}=\left|\frac{\lambda_{c}^{(s)}}{\lambda_{c}^{(s)}}\right|
\frac{A_{c}}{|A_{c}|}
\end{equation}
\begin{equation}
\epsilon_{A} e^{i\phi_{A}}=\left|\frac{\lambda_{u}^{(s)}}{\lambda_{c}^{(s)}}\right|\frac{A_{u}}{|A_{c}|}
\end{equation}
we reobtain expression (\ref{rate}):
\begin{equation}
R=\frac{\left|\epsilon e^{i\phi_{T}}(e^{i\gamma}-\delta_{EW})
-\epsilon_{A} e^{i\phi_{A}}e^{i\gamma}-e^{i\phi_{P}}\right|^{2}
+\left|\epsilon e^{i\phi_{T}}(e^{-i\gamma}-\delta_{EW})
-\epsilon_{A} e^{i\phi_{A}}e^{-i\gamma}-e^{i\phi_{P}}\right|^{2}}
{\left|\epsilon_{A} e^{i\phi_{A}}e^{i\gamma}+e^{i\phi_{P}}\right|^{2}+
\left|\epsilon_{A} e^{i\phi_{A}}e^{-i\gamma}+e^{i\phi_{P}}\right|^{2}}
\end{equation}
and the constraint:
\begin{equation}
|\cos \gamma - \delta_{EW}|\geq
\frac{|1-R|}{2\epsilon},
\end{equation}
where the parameter $\epsilon$ is, again, 
\begin{equation}
\epsilon=\sqrt{2}\,\frac{f_{K}}{f_{\pi}}\left|\frac{{\cal A}(B^{+}\rightarrow\pi^{0}\pi^{+})}{{\cal A}(B^{+} \rightarrow K^{0} \pi^{+})}\right|
\left|\frac{V_{us}}{V_{ud}}\right|.
\end{equation}
\subsection{The Neubert-Rosner Method in \mbox{\boldmath$B \rightarrow PP $} Decays}
In order to try to find similar constraints in $B \rightarrow PP$ decays, all that we need to do is to look again at the tables of appendix A and look for relations of the form of equation (\ref{nram}). We immediately find:
\begin{itemize}
\item For $\Delta S=1$ decays
\begin{equation}
 \sqrt{2}{\cal A}(B^{0} \rightarrow K^{0} \pi^{0})-
{\cal A}(B^{0} \rightarrow K^{+} \pi^{-})=20 a'_{5}
\end{equation}
\item For $\Delta S=0$ decays:
\begin{eqnarray}
\sqrt{2}{\cal A}(B^{0}_{s} \rightarrow K^{0} \pi^{0})-
{\cal A}(B^{0}_{s} \rightarrow K^{-} \pi^{+})&=&20 a_{5} \\
\sqrt{2}{\cal A}(B^{0} \rightarrow \pi^{0} \pi^{0})-
{\cal A}(B^{0} \rightarrow \pi^{-} \pi^{+})&=&20 a_{5}.
\end{eqnarray}
\end{itemize}
Let us first look at the $\Delta S=1$ process. We can define:
 \begin{eqnarray}
\sqrt{2}{\cal A}(B^{+} \rightarrow \pi^{0} \pi^{+})
&=& +|\lambda_{u}^{(d)}|e^{i\gamma}B \nonumber \\
\sqrt{2}{\cal A}(B^{0} \rightarrow K^{0} \pi^{0})
&=& +\lambda_{u}^{(s)}A_{u}+\lambda_{c}^{(s)}A_{c}+
|\lambda_{u}^{(s)}|(e^{i\gamma}-\delta_{EW})B \\
{\cal A}(B^{0} \rightarrow K^{+} \pi^{-})
&=& +\lambda_{u}^{(s)}A_{u}+\lambda_{c}^{(s)}A_{c}\nonumber 
\end{eqnarray}
where in here:
\begin{eqnarray}
B&=&-20\cdot\frac{1}{2}(c_{2}+c_{1})
\sqrt{\frac{1}{200}}\langle{27}\|\overline{15}\|3\rangle\nonumber \\  
A_{u}&=&-3\cdot a_{u}\sqrt{\frac{1}{15}}\langle{8}_{S}\|\overline{3}\|3\rangle) \nonumber \\
&&-\frac{1}{2}(c_{2}-c_{1})\sqrt{\frac{1}{5}}\langle{8}_{S}\|6\|3\rangle
\nonumber\\
&&+\frac{1}{2}(c_{2}+c_{1})\sqrt{\frac{1}{50}}\langle{8}_{S}\|\overline{15}\|3\rangle\nonumber\\
&&+8\cdot\frac{1}{2}(c_{2}+c_{1})\sqrt{\frac{1}{200}}\langle{27}\|\overline{15}\|3\rangle\nonumber\\
A_{c}&=&
-3\cdot a_{c}\sqrt{\frac{1}{15}}\langle{8}_{S}\|\overline{3}\|3\rangle \nonumber \\
&&-\frac{1}{2}\frac{|\lambda^{(s)}_{u}|\delta_{EW}}{\lambda^{(s)}_{c}}(c_{2}-c_{1})\sqrt{\frac{1}{5}}\langle{8}_{S}\|6\|3\rangle
\nonumber\\
&&+\frac{1}{2}\frac{|\lambda^{(s)}_{u}|\delta_{EW}}{\lambda^{(s)}_{c}}(c_{2}+c_{1})\sqrt{\frac{1}{50}}\langle{8}_{S}\|\overline{15}\|3\rangle\nonumber\\
&&+8\cdot\frac{1}{2}\frac{|\lambda^{(s)}_{u}|\delta_{EW}}{\lambda^{(s)}_{c}}(c_{2}+c_{1})\sqrt{\frac{1}{200}}\langle{27}\|\overline{15}\|3\rangle.
\end{eqnarray}
We can repeat the previous procedure, which was applied to 
$B^{+} \rightarrow K\pi$. We immediately obtain a new constraint:
\begin{equation}
\label{newconst}
|\cos \gamma - \delta_{EW}|\geq
\frac{|1-R^{*}|}{2\epsilon^{*}},
\end{equation}
where:
\begin{equation}
R^{*}=\frac{2[Br(B^{0} \rightarrow K^{0} \pi^{0})+
Br(\overline{B^{0}} \rightarrow \overline{K^{0}} \pi^{0})]}
{Br(B^{0} \rightarrow K^{+} \pi^{-})+
Br(\overline{B^{0}} \rightarrow K^{-} \pi^{+})}
\end{equation}
and 
\begin{equation}
\epsilon^{*}=\sqrt{2}\,\frac{f_{K}}{f_{\pi}}\left|\frac{{\cal A}(B^{+}\rightarrow\pi^{0}\pi^{+})}{{\cal A}(B^{0} \rightarrow K^{+} \pi^{-})}\right|
\left|\frac{V_{us}}{V_{ud}}\right|.
\end{equation}\\
 From the current data the value of $\epsilon^{*}$ is \cite{pdg}: $0.09\pm 0.04$ \cite{babar1} (BaBar), $0.13\pm 0.06$ \cite{belle1} (Belle), $0.25\pm 0.19$ \cite{cassel} (CLEO). 

We now turn to the $\Delta S=0$ relations. We can, in principle, obtain constraints in a similar fashion, However, trying to repeat the procedure we immediately encounter a problem. In $\Delta S=1$ decays we have a ``small'' parameter, which is \cite{pdg}:
\begin{equation}
0.01\leq\left|\frac{\lambda^{(s)}_{u}}{\lambda^{(s)}_{c}}\right|\leq0.03
\end{equation}
In $\Delta S=0$ the equivalent ratio would be \cite{pdg}:
\begin{equation}
0.2\leq\left|\frac{\lambda^{(d)}_{u}}{\lambda^{(d)}_{c}}\right|\leq 0.6
\end{equation}
which is much larger. Since the expansion parameter $\tilde{\epsilon}$ is proportional to this ratio, it is doubtful whether $\tilde{\epsilon}$ would be smaller then $1$. In fact from the current data we find that for $B^{0} \rightarrow \pi\pi$ decays\cite{babar1,belle1,cassel}:
\begin{equation}
\tilde{\epsilon}=\sqrt{2}\left|\frac{{\cal A}(B^{+}\rightarrow\pi^{0}\pi^{+})}{{\cal A}(B^{0} \rightarrow \pi^{-} \pi^{+})}\right| > 1.
\end{equation} 
( notice that we didn't need the $SU(3)$ breaking factor since we don't need to relate $\Delta S=1$ processes to $\Delta S=0$ processes).
We see that the expansion in $\tilde{\epsilon}$ does not make sense and no constraint can be obtained.\\ 
 Similarly, it is doubtful whether the expansion would be justified when discussing $B_{s} \rightarrow K\pi$ decays. Therefore, apart from the original Neubert-Rosner constraint, the only constraint we can obtain in $B \rightarrow PP$ decays is (\ref{newconst}).  

\subsection{The Neubert-Rosner Method in \mbox{\boldmath$B \rightarrow VP $} Decays}We now turn to discuss $B\rightarrow VP$ decays.\\ 
In $B\rightarrow PP$ there are four pairs of processes of the form 
first investigated by Neubert and Rosner:
\begin{itemize}
\item $B^{+} \rightarrow K\pi$ (the original Neubert-Rosner pair)
\item $B^{0} \rightarrow K\pi$
\item $B^{0}_{s} \rightarrow K\pi$
\item $B^{0} \rightarrow \pi\pi$
\end{itemize}
We would expect to find eight pairs of such processes since, 
as a rule of thumb, in $B\rightarrow VP$  there are twice 
as much decays. In fact using the tables of Appendix A we can  
find only six pairs, which are:
\begin{itemize}
\item $\Delta S =1$ decays
\begin{eqnarray}
\sqrt{2}{\cal A}(B^{+}\rightarrow \rho ^{0} K^{+} )
+{\cal A}(B^{+}\rightarrow \rho ^{+} K^{0}) & = &
20b'_{5}+6b'_{8}  \label{rhok} \\
\sqrt{2}{\cal A}(B^{+}\rightarrow K^{*+} \pi ^{0})
+{\cal A}(B^{+}\rightarrow  K^{*0} \pi^{+}) & = &
20b'_{5}-6b'_{8}  \\
&&  \nonumber\\
\sqrt{2}{\cal A}(B^{0}\rightarrow \rho ^{0} K^{0})
-{\cal A}(B^{0}\rightarrow \rho ^{-} K^{+}) & = &
20b'_{5}+6b'_{8} \label{rhok0} \\
\sqrt{2}{\cal A}(B^{0}\rightarrow K^{*0} \pi ^{0})
-{\cal A}(B^{0}\rightarrow  K^{*+} \pi^{-}) & = &
20b'_{5}-6b'_{8} \label{k*pi0} 
\end{eqnarray}
\item $\Delta S =0$ decays
\begin{eqnarray}
\sqrt{2}{\cal A}(B^{0}_{s}\rightarrow \rho ^{0} \overline{K^{0}})
-{\cal A}(B^{0}_{s}\rightarrow \rho ^{+} K^{-}) & = &
20b_{5}+6b_{10}  \label{DS0a}\\
\sqrt{2}{\cal A}(B^{0}_{s}\rightarrow \overline{K^{*0}} \pi ^{0})
-{\cal A}(B^{0}_{s}\rightarrow  K^{*-} \pi^{+}) & = &
20b_{5}-6b_{10}  \label{DS0b}.
\end{eqnarray}
\end{itemize}  
We note that the differences between the amplitudes have:
\begin{itemize}
\item 2 weak phases $e^{i\gamma}-\delta_{EW}$, $e^{i\gamma}+\delta_{EW}$
in  $\Delta S =1$ decays
\item 1 weak phase $e^{i\gamma}-\bar{\delta}_{EW}$ in  $\Delta S =0$ decays.
\end{itemize}
Unfortunately we cannot use {\bf any} of these pairs because the differences
are {\bf unmeasurable} through some decay. Stated differently, we don't have the equivalent of $B^{+}\rightarrow \pi^{0} \pi^{+}$ in $B\rightarrow VP$ decays. the ``natural'' candidates, namely, $B^{+}\rightarrow \rho^{0} \pi^{+}$
or $B^{+}\rightarrow \rho^{+} \pi^{0}$ cannot be related to these differences
as can be easily seen from table 7. Thus we come to the conclusion that the Neubert-Rosner method cannot be used in $B\rightarrow VP$ without further assumptions. 
\subsection{Neubert-Rosner Method in \mbox{\boldmath$B \rightarrow VP $} Decays \\
 with Further Assumptions}
The assumption that we need is to neglect some parts of the amplitudes. This assumption goes beyond pure $SU(3)$ symmetry and it can only be made using a graphical analysis. In a previous article, Gronau was able to obtain a constraint on $\gamma$ based on this assumption \cite{vp}. We shall now present this assumption and relate it to our analysis.\\
In $B\rightarrow VP$ decays we have 12 different graphical amplitudes \cite{vp}: $T_{M}$, $C_{M}$, $P_{M}$, $E_{M}$, $A_{M}$, $PA_{M}$. The suffix $M=P,V$ for $T_{M}$, $C_{M}$, $P_{M}$ denotes whether the spectator quark is included in a pseudoscalar or vector meson. In $E_{M}$, $A_{M}$, $PA_{M}$ it denotes the type of meson into which the outgoing quark $q_{3}$ enters in $\overline{b}q_{1}\rightarrow \overline{q}_{2} q_{3}$.\\
In $\Delta S = 0$ decays one can safely neglect the EWP contribution  ( recall that $\bar{\delta}_{EW}=0.03$ ), doing this, one finds that: 
\begin{table}[h]
\begin{center}
\begin{tabular}{lcc}
\hline\hline
Decay mode && Tree amplitude \\ \hline\\
$B^{+}\rightarrow \rho^{+} \pi^{0}$ &&
$-\frac{1}{\sqrt{2}}(T_{P}+C_{V}+P_{u,P}-P_{u,V}+A_{P}-A_{V})$ \\\\
$B^{+}\rightarrow \rho^{0} \pi^{+}$ &&
$-\frac{1}{\sqrt{2}}(T_{V}+C_{P}+P_{u,V}-P_{u,P}+A_{V}-A_{P})$ \\\\
$B^{+}\rightarrow K^{*+} \overline{K^{0}}$ &&
$A_{V}+P_{u,V}$ \\\\
$B^{+}\rightarrow \overline{K^{*0}} K^{+}$ &&
$A_{P}+P_{u,P}$ \\\hline\hline \\\\
\end{tabular}
\caption{Graphical Analysis of $B\rightarrow VP$, $\Delta S=0$ decays (partial list) }
\end{center}
\end{table}
\clearpage
In $\Delta S = 1$ decays, by separating the EWP contribution from the Tree contribution, one finds that:

\begin{table}[h]
\begin{center}
\begin{tabular}{lcccc}
\hline\hline
Decay mode && Tree amplitude && EWP amplitude \\ \hline\\
$B^{+}\rightarrow \rho^{+} K^{0}$ &&
$A_{V}+P_{u,V}$ && $\frac{\kappa}{2}(C_{V}-2E_{V}+P_{u,V})$\\\\
$B^{+}\rightarrow \rho^{0} K^{+}$ &&
$-\frac{1}{\sqrt{2}}(T_{V}+C_{P}+A_{V}+P_{u,V})$ &&
$\frac{\kappa}{2\sqrt{2}}(3T_{P}+2C_{V}+2E_{V}-P_{u,V})$\\\\
$B^{0}\rightarrow \rho^{-} K^{+}$ &&
$-(T_{V}+P_{u,V})$ && $\frac{\kappa}{2}(2C_{V}-E_{V}-P_{u,V})$\\\\
$B^{0}\rightarrow \rho^{0} K^{0}$ &&
$\frac{1}{\sqrt{2}}(-C_{P}+P_{u,V})$ &&
$\frac{\kappa}{2\sqrt{2}}(3T_{P}+C_{V}+E_{V}+P_{u,V})$ \\\\
\\\hline\hline
\end{tabular}
\caption{Graphical Analysis of $B\rightarrow VP$, $\Delta S=1$ decays (partial list)}
\end{center}
\end{table}

Let us look at the $B^{+}\rightarrow\rho K $ pairs. Adding the amplitudes one finds that:
\begin{equation}
\sqrt{2}{\cal A}(B^{+}\rightarrow \rho^{0} K^{+})+
{\cal A}(B^{+}\rightarrow \rho^{+} K^{0})=
-(T_{V}+C_{P})+\frac{3}{2}\kappa(T_{P}+C_{V})
\end{equation}
(compare this to equation (\ref{rhok})). Looking now at the $\Delta S = 0$ decays we see that although $B^{+}\rightarrow\rho^{+}\pi^{0}$ contains $T_{P}+C_{V}$, it also contains $P_{u,M}$ and $A_{M}$ diagrams. Similarly, $B^{+}\rightarrow\rho^{0}\pi^{+}$ contains $T_{V}+C_{P}$ and also $P_{u,M}$ and $A_{M}$. If one neglects them with respect to $T$ and $C$ diagrams (since $P_{u,M}$ and $A_{M}$ represent rescattering contribution), a constraint can be obtained \cite{vp}.\\
We can reformulate this assumption using $B^{+} \rightarrow K^{*}K $ decays. Since :
\begin{eqnarray}
-(T_{P}+C_{V}) & = & \sqrt{2}{\cal A}(B^{+}\rightarrow\rho^{+}\pi^{0})
+{\cal A}(B^{+} \rightarrow \overline{K^{*0}}K^{+})
 -{\cal A}(B^{+} \rightarrow K^{*+}\overline{K^{0}})\nonumber \\
-(T_{V}+C_{P}) & = & \sqrt{2}{\cal A}(B^{+}\rightarrow\rho^{0}\pi^{+})
+{\cal A}(B^{+} \rightarrow K^{*+}\overline{K^{0}})            
-{\cal A}(B^{+} \rightarrow \overline{K^{*0}}K^{+}),\\
\end{eqnarray}
neglecting the rescattering contributions leads to:  
\begin{eqnarray}
-(T_{P}+C_{V}) &\approx& \sqrt{2}{\cal A}(B^{+}\rightarrow\rho^{+}\pi^{0})
\nonumber \\
-(T_{V}+C_{P}) &\approx& \sqrt{2}{\cal A}(B^{+}\rightarrow\rho^{0}\pi^{+}).
\end{eqnarray}
It is clear that the assumption is equivalent to the statement that 
\begin{eqnarray}
\label{aprox}
\sqrt{2}{\cal A}(B^{+}\rightarrow\rho^{+}\pi^{0}) 
\approx\sqrt{2}{\cal A}(B^{+}\rightarrow\rho^{+}\pi^{0})
+{\cal A}(B^{+} \rightarrow \overline{K^{*0}}K^{+})
 -{\cal A}(B^{+} \rightarrow K^{*+}\overline{K^{0}})\nonumber \\
\sqrt{2}{\cal A}(B^{+}\rightarrow\rho^{0}\pi^{+})\approx
\sqrt{2}{\cal A}(B^{+}\rightarrow\rho^{0}\pi^{+})
+{\cal A}(B^{+} \rightarrow K^{*+}\overline{K^{0}})            
-{\cal A}(B^{+} \rightarrow \overline{K^{*0}}K^{+}).
\end{eqnarray}
Using table 7 of Appendix A, we see that the previous approximation can be written as:
\begin{eqnarray}
\sqrt{2}{\cal A}(B^{+}\rightarrow\rho^{+}\pi^{0})&\approx&
20b_{5}-6b_{8} \nonumber \\
\sqrt{2}{\cal A}(B^{+}\rightarrow\rho^{0}\pi^{+})&\approx&
20b_{5}+6b_{8}. 
\end{eqnarray}
In a more explicit way, by neglecting $\bar{\delta}_{EW}$, we can write:
\begin{eqnarray}
20b_{5}-6b_{8}=-\frac{1}{2}|\lambda^{(d)}_{u}|e^{i\gamma}
\left[(c_{2}+c_{1})\langle{27}\|\overline{15}\|3\rangle
-(c_{2}-c_{1})\langle10\|6\|3\rangle\right] & \equiv &
|\lambda^{(d)}_{u}|e^{i\gamma} B_{-}  \nonumber  \\
20b_{5}+6b_{8}=-\frac{1}{2}|\lambda^{(d)}_{u}|e^{i\gamma}
\left[(c_{2}+c_{1})\langle{27}\|\overline{15}\|3\rangle
+(c_{2}-c_{1})\langle10\|6\|3\rangle\right] & \equiv &
|\lambda^{(d)}_{u}|e^{i\gamma} B_{+},
\end{eqnarray}
so eventually:
\begin{eqnarray}
\sqrt{2}{\cal A}(B^{+}\rightarrow\rho^{+}\pi^{0})&\approx&
|\lambda^{(d)}_{u}|e^{i\gamma} B_{-}  \nonumber \nonumber \\
\sqrt{2}{\cal A}(B^{+}\rightarrow\rho^{0}\pi^{+})&\approx&
|\lambda^{(d)}_{u}|e^{i\gamma} B_{+}. 
\end{eqnarray}
(notice that $B_{-}$ and $B_{+}$ are complex numbers).\\
We can now return to equation (\ref{rhok}), which can be written as:
\begin{equation}
-{\cal A}(B^{+}\rightarrow \rho ^{+} K^{0})+20b'_{5}+6b'_{8}
=\sqrt{2}{\cal A}(B^{+}\rightarrow \rho ^{0} K^{+}).
\end{equation}
If we define 
\begin{equation}
{\cal A}(B^{+}\rightarrow \rho ^{+} K^{0})=
\lambda_{u}^{(s)}A_{u}+\lambda_{c}^{(s)}A_{c},
\end{equation}
we can write:
\begin{equation}
\sqrt{2}{\cal A}(B^{+}\rightarrow \rho ^{0} K^{+})=
-\lambda_{u}^{(s)}A_{u}-\lambda_{c}^{(s)}A_{c}
+|\lambda^{(s)}_{u}|e^{i\gamma}B_{+}
-|\lambda^{(s)}_{u}|\delta_{EW}B_{-}
\end{equation}
since 
\begin{equation}
20b'_{5}+6b'_{8}=|\lambda^{(s)}_{u}|e^{i\gamma}B_{+}
-|\lambda^{(s)}_{u}|\delta_{EW}B_{-}.
\end{equation} 
We now turn  to the CP averaged ratio:
\begin{equation}
R^{+}_{\rho k}=\frac{2[Br(B^{+} \rightarrow  \rho^{0} K^{+})+
Br(B^{-} \rightarrow \rho^{0} K^{-})]}
{Br(B^{+} \rightarrow \rho^{+} K^{0})+
Br(B^{-} \rightarrow \rho^{-} \overline{K^{0}})}, 
\end{equation} 
and define: 
\begin{eqnarray}
e^{i\phi_{0}}=\left|\frac{\lambda_{u}^{(s)}}{\lambda_{c}^{(s)}}\right|
\frac{A_{c}}{|A_{c}|} & , &
\epsilon_{A}e^{i\phi_{A}}=\left|\frac{\lambda_{u}^{(s)}}{\lambda_{c}^{(s)}}\right|\frac{A_{u}}{|A_{c}|} \nonumber  \\ \nonumber\\
\epsilon_{V}e^{i\phi_{V}}=\left|\frac{\lambda_{u}^{(s)}}{\lambda_{c}^{(s)}}\right|\frac{B_{+}}{|A_{c}|} & , &
\epsilon_{P}e^{i\phi_{P}}=\left|\frac{\lambda_{u}^{(s)}}{\lambda_{c}^{(s)}}\right|\frac{B_{-}}{|A_{c}|},
\end{eqnarray}
so the ratio is:
\begin{equation}
R^{+}_{\rho k}=\frac{\left|-e^{i\phi_{0}}-\epsilon_{A}e^{i\phi_{A}}e^{i\gamma}
+\epsilon_{V}e^{i\phi_{V}}e^{i\gamma}-\delta_{EW}\epsilon_{P}e^{i\phi_{P}}
e^{i\gamma}\right|^{2}+ CP}
{\left|-e^{i\phi_{0}}-\epsilon_{A}e^{i\phi_{A}}e^{i\gamma}\right|^{2}+ CP}.
\end{equation}
Expanding, we find that in the lowest order:
\begin{equation}
R^{+}_{\rho k}\approx 1- 2\epsilon_{V}\cos(\phi_{V}-\phi_{0})\cos{\gamma}
+2\delta_{EW}\epsilon_{P}\cos(\phi_{P}-\phi_{0}),
\end{equation}
which leads to the constraint\cite{vp}:
\begin{equation}
|\cos{\gamma}|\geq \frac{|1-R^{+}_{\rho k}|}{2\epsilon_{V}}-
\delta_{EW}\left(\frac{\epsilon_{P}}{\epsilon_{V}}\right).
\end{equation}
The parameters $\epsilon_{V}$, $\epsilon_{P}$
are related to measured quantities by:
\begin{eqnarray}
\epsilon_{V}&=&\sqrt{2}\,\frac{f_{K}}{f_{\pi}}\left|\frac{{\cal A}(B^{+}\rightarrow\rho^{0}\pi^{+} )}{{\cal A}(B^{+} \rightarrow\rho^{+} K^{0})}\right|
\frac{V_{us}}{V_{ud}} \nonumber \\
\epsilon_{P}&=&\sqrt{2}\,\frac{f_{K^{*}}}{f_{\rho}}\left|\frac{{\cal A}(B^{+}\rightarrow\rho^{+}\pi^{0})}{{\cal A}(B^{+} \rightarrow\rho^{+} K^{0})}\right|
\frac{V_{us}}{V_{ud}}.
\end{eqnarray}
The various decay constants appear as a result of factorization, which was used to relate $B_{-}\,(= T_{V}+C_{P})$ and $B_{+}\,(= T_{P}+C_{V})$ of $\Delta S=1$ decays to $B_{-}$ and $B_{+}$ of $\Delta S=1$ decays. In relating $B_{+}$ we are ``exchanging'' $\pi$ and $K$ (exchange of a pseudoscalar). In relating $B_{-}$ we are ``exchanging'' $\rho$ and $K^{*}$ (exchange of a vector).\\
We see that a weaker constraint can be obtained based on the additional assumption. The reason that it is weaker is the fact that there are two weak phases in the difference of the amplitudes. \\
A similar constraint can be obtained using
$B \rightarrow {K^{*}\pi}$ decays. Since 
\begin{equation}
20b'_{5}-6b'_{8}=|\lambda^{(s)}_{u}|e^{i\gamma}B_{-}
-|\lambda^{(s)}_{u}|\delta_{EW}B_{+},
\end{equation} 
one can repeat the procedure using the CP averaged ratio:
\begin{equation}
R^{+}_{k^{*}\pi}=\frac{2[Br(B^{+} \rightarrow K^{*+}\pi^{0})+
Br(B^{-} \rightarrow K^{*-}\pi^{0} )]}
{Br(B^{+} \rightarrow K^{*0}\pi^{+})+
Br(B^{-} \rightarrow \overline{K^{*0}}\pi^{-})} 
\end{equation} 
and obtain a constraint\cite{vp}:
\begin{equation}
|\cos{\gamma}|\geq \frac{|1-R^{+}_{k^{*} \pi}|}{2\epsilon^{*}_{P}}-
\delta_{EW}\left(\frac{\epsilon^{*}_{V}}{\epsilon^{*}_{P}}\right),
\end{equation}
where 
\begin{eqnarray}
\epsilon^{*}_{V}&=&\sqrt{2}\,\frac{f_{K}}{f_{\pi}}\left|\frac{{\cal A}(B^{+}\rightarrow\rho^{0}\pi^{+})}{{\cal A}(B^{+} \rightarrow K^{*0}\pi^{+} )}\right|
\frac{V_{us}}{V_{ud}} \nonumber \\
\epsilon^{*}_{P}&=&\sqrt{2}\,\frac{f_{K^{*}}}{f_{\rho}}\left|\frac{{\cal A}(B^{+}\rightarrow\rho^{+}\pi^{0})}{{\cal A}(B^{+} \rightarrow K^{*0}\pi^{+} )}\right|
\frac{V_{us}}{V_{ud}}. 
\end{eqnarray}

We now turn to relations (\ref{rhok0}) and (\ref{k*pi0}).\\
Following the same line of reasoning we find,  based on the expansion of the ratios
\begin{eqnarray}
R^{0}_{\rho k}&=&\frac{2[Br(B^{0} \rightarrow  \rho^{0} K^{0})+
Br(\overline{B^{0}} \rightarrow \rho^{0} \overline{K^{0}})]}
{Br(B^{0} \rightarrow \rho^{-} K^{+})+Br(\overline{B^{0}} \rightarrow \rho^{+} K^{-})} \\ \nonumber \\
R^{0}_{k^{*}\pi}&=&\frac{2[Br(B^{0} \rightarrow K^{*0}\pi^{0})+
Br(\overline{B^{0}} \rightarrow \overline{K^{*0}}\pi^{0} )]}
{Br(B^{0} \rightarrow K^{*+}\pi^{-})+Br(\overline{B^{0}} \rightarrow K^{*-}\pi^{+})}, 
\end{eqnarray}
two new constraints:
\begin{eqnarray}
|\cos{\gamma}|&\geq& \frac{|1-R^{0}_{\rho k}|}{2\tilde{\epsilon}_{V}}-
\delta_{EW}\left(\frac{\tilde{\epsilon}_{P}}{\tilde{\epsilon}_{V}}\right). \\ \nonumber \\
|\cos{\gamma}|&\geq& \frac{|1-R^{0}_{k^{*} \pi}|}{2\tilde{\epsilon}^{*}_{P}}-
\delta_{EW}\left(\frac{\tilde{\epsilon}^{*}_{V}}{\tilde{\epsilon}^{*}_{P}}\right), 
\end{eqnarray}
where 
\begin{eqnarray}
\tilde{\epsilon}_{V}&=&\sqrt{2}\,\frac{f_{K}}{f_{\pi}}\left|\frac{{\cal A}(B^{+}\rightarrow\rho^{0}\pi^{+})}{{\cal A}(B^{0} \rightarrow\rho^{-} K^{+})}\right|
\frac{V_{us}}{V_{ud}} \nonumber \\
\tilde{\epsilon}_{P}&=&\sqrt{2}\,\frac{f_{K^{*}}}{f_{\rho}}\left|\frac{{\cal A}(B^{+}\rightarrow\rho^{+}\pi^{0})}{{\cal A}(B^{0} \rightarrow\rho^{-} K^{+})}\right|\frac{V_{us}}{V_{ud}} \nonumber \\
\tilde{\epsilon}^{*}_{V}&=&\sqrt{2}\,\frac{f_{K}}{f_{\pi}}\left|\frac{{\cal A}(B^{+}\rightarrow\rho^{0}\pi^{+})}{{\cal A}(B^{0} \rightarrow K^{*+}\pi^{-} )}\right|\frac{V_{us}}{V_{ud}} \nonumber \\
\tilde{\epsilon}^{*}_{P}&=&\sqrt{2}\,\frac{f_{K^{*}}}{f_{\rho}}\left|\frac{{\cal A}(B^{+}\rightarrow\rho^{+}\pi^{0})}{{\cal A}(B^{0} \rightarrow K^{*+}\pi^{-} )}\right|\frac{V_{us}}{V_{ud}}.
\end{eqnarray}

Finally we will consider the $\Delta S =0 $ relations (\ref{DS0a}) and (\ref{DS0b}). Using the graphical method we immediately find (neglecting as usual the EWP contribution in $\Delta S =0 $ decays):\\\\
\begin{tabular}{lrll}
I&$\sqrt{2}{\cal A}(B^{0}_{s}\rightarrow \rho ^{0} \overline{K^{0}})$
&$=$&$- C_{P}+P_{u,P}$ \\ \\
II& ${\cal A}(B^{0}_{s}\rightarrow \rho ^{+} K^{-})$ 
&$=$& $ - T_{P}-P_{u,P}$ \\\\
III&$\sqrt{2}{\cal A}(B^{0}_{s}\rightarrow \overline{K^{*0}} \pi ^{0})$
&$=$& $- C_{V}+P_{u,V}$ \\\\
IV&${\cal A}(B^{0}_{s}\rightarrow  K^{*-} \pi^{+})$
&$=$ & $- T_{V}-P_{u,V}$.
\end{tabular}\\

There is a phase convention difference between the graphical method and our analysis, therefore we need to add I and II (or III and IV) rather then subtract them. Adding up we find:
\begin{eqnarray}
\sqrt{2}{\cal A}(B^{0}_{s}\rightarrow \rho ^{0} \overline{K^{0}})
+{\cal A}(B^{0}_{s}\rightarrow \rho ^{+} K^{-}) & = &  -(T_{P}+C_{P})\\
\sqrt{2}{\cal A}(B^{0}_{s}\rightarrow \overline{K^{*0}} \pi ^{0})
+{\cal A}(B^{0}_{s}\rightarrow  K^{*-} \pi^{+}) & = &  -(T_{V}+C_{V}).
\end{eqnarray}
These differences are {\bf unmeasurable} even by neglecting some parts of the amplitudes because there is no decay mode which contains $T_{P}+C_{P}$ or 
$T_{V}+C_{V}$ as we now explain.\\
A decay which contain both $T$ and $C$ must be a $B^{+}$ decay.
Furthermore, if $T_{V}$ ($T_{P}$) contributes then the vector (pseudoscalar) meson must contain the combination $\overline{u}u$. If $C_{V}$ ($C_{P}$) contributes then this meson must contain the combination $\overline{q}u$ where ($q=d,s$). Obviously all theses demands cannot be satisfied simultaneously. Therefore 
$T_{P}+C_{P}$ or $T_{V}+C_{V}$ cannot be measured through some decay. \\
In fact, even if we were able to measure it, it is doubtful that a useful constraint could have been obtained. The reason is that the expansion parameter would have to contain the ratio $\left|\frac{\lambda^{(d)}_{u}}{\lambda^{(d)}_{c}}\right|$ and as we saw before, it might make the expansion parameter too large. Anyway, no constraint can be obtained from the $\Delta S=0 $ decays into a $VP$ pairs.

\section{Discussion and Conclusions}
When we presented the Neubert-Rosner method in section 2 we noted that there are four essential features (or even demands) to this method. They are:
\begin{enumerate}
\item The EWP contribution ($Q_{9}$,$Q_{10}$) can be related to the Tree contribution.
\item There are two amplitudes that differ in a complex number, which has one strong phase and a unique weak phase.
\item This number can be measured through some decays.
\item The difference between the amplitudes is small compared to (one of) the amplitudes. 
\end{enumerate}
As we have seen the first demand was very easy to fulfill and in fact we were able (in section 2) to relate the EWP contribution to the Tree contribution in a systematic way. \\
When we discussed $B \rightarrow PP$ decays we saw that 
in $\Delta S =1 $ decays we were able to fulfill all of the remaining demands.
In $\Delta S =0 $ decays we saw that the fourth demand probably cannot be fulfilled and therefore the constraints, that can be obtained, are incorrect.\\
In $B \rightarrow VP$ decays we saw that all of the possible ``candidates'' 
were not suitable for constraints because the third demand could not be fulfilled without further assumptions. Making those assumptions, i.e., neglecting some parts of the amplitudes, some constraints can be obtained although they are manifestly weaker. \\
The reason for this ``weakening'' is the fact that in $\Delta S=1 $ decays {\bf 2} different weak phases are involved (in contrast with the second demand). In $\Delta S=0 $ decays we weren't able to find constraints even after neglecting some parts of the amplitudes.\\
These remarks are summarized in table \ref{sumtable} (3' denotes the third demand after neglecting some parts of the amplitudes).
\begin{table}[h]
\label{sumtable}
\begin{tabular}{|c|c|c|c|c|c|l|c|}\hline
Decay mode&Dem.&Dem.&Dem.&Dem.&Dem.&Constraint&Remarks \\
&1&2&3&3'&4&made with&\\  \hline
$B \rightarrow PP$  &+&+&+&&+&$B^{+} \rightarrow K\pi$
&Neubert-Rosner constraint\\ \cline{2-8}
$\Delta S=0$&+&+&+&&+&$B^{0} \rightarrow K\pi$
&new constraint\\ \hline
$B \rightarrow PP$ &+&+&+&&$-$&\hspace{5 mm}none&\\ 
$\Delta S=0$&&&&&&&\\ \hline
$B \rightarrow VP$ &+&$-$&$-$&+&+&$B^{+} \rightarrow \rho K$
&Gronau constraint, weaker\\ \cline{2-8}
$\Delta S=1$&+&$-$&$-$&+&+&$B^{+} \rightarrow K^{*}\pi$&Gronau constraint, weaker\\ \cline{2-8}
&+&$-$&$-$&+&+&$B^{0} \rightarrow \rho K$
&new constraint, weaker\\ \cline{2-8}
&+&$-$&$-$&+&+&$B^{0} \rightarrow K^{*}\pi$&new constraint, weaker\\ \hline
$B \rightarrow VP$ &+&+&$-$&$-$&&\hspace{5 mm}none&\\ 
$\Delta S=0$&&&&&&&\\ \hline
\end{tabular}
\caption{Summary}
\end{table}
It is quite obvious that the reason that the Neubert-Rosner Method was almost inapplicable in $B \rightarrow VP$ decays is the fact that the final states are not symmetrical. It was this property which allowed us to connect the differences of pairs of amplitudes to measurable quantities. Stated differently, without symmetrization the decay product $\;\pi^{0}\pi^{+}\;$ is not pure 
$\;I=\frac{3}{2}\;$ state and therefore contains ``too many''  reduced matrix elements.\\
An essential feature of the $B \rightarrow VP$ decays is the large number of of decay products. Roughly speaking for each decays product in the $PP$ system we can find two analogous decay products in the $VP$ system, namely, $VP$ and $PV$ (see section 3 for examples). A notable exception to this ``rule'' is the decay product $\rho^{0} \pi^{0}$ (the equivalent of $\pi^{0}\pi^{0}$) which is already symmetrical. The natural question arises ``can we use this feature?'', or stated differently, ``can we turn the weakness into strength?''.
\newpage
Using  the tables of appendix A we can write:
\begin{eqnarray}
{\cal A}(B^{0}_{s} \rightarrow \rho^{+}\pi^{-})&=&
\lambda_{c}^{(s)}S_{c}+\lambda_{u}^{(s)}S_{u}+
\lambda_{c}^{(s)}A_{c}+\lambda_{u}^{(s)}A_{u}. \nonumber \\
{\cal A}(B^{0}_{s} \rightarrow \rho^{-}\pi^{+})&=&
\lambda_{c}^{(s)}S_{c}+\lambda_{u}^{(s)}S_{u}-
\lambda_{c}^{(s)}A_{c}-\lambda_{u}^{(s)}A_{u}. \nonumber \\
{\cal A}(B^{0}_{s} \rightarrow \rho^{0}\pi^{0})&=&
\lambda_{c}^{(s)}S_{c}+\lambda_{u}^{(s)}S_{u}. \nonumber \\
{\cal A}(B^{0} \rightarrow K^{*+}K^{-})&=&
\lambda_{c}^{(d)}S_{c}+\lambda_{u}^{(d)}S_{u}+
\lambda_{c}^{(d)}A_{c}+\lambda_{u}^{(d)}A_{u}. \nonumber \\
{\cal A}(B^{0} \rightarrow K^{*-}K^{+})&=&
\lambda_{c}^{(d)}S_{c}+\lambda_{u}^{(d)}S_{u}-
\lambda_{c}^{(d)}A_{c}-\lambda_{u}^{(d)}A_{u}. \nonumber \\
\end{eqnarray}
Where $S$ ($A$) denotes the symmetric (anti-symmetric) part of the amplitude under an exchange of the $SU(3)$ quantum numbers of the (pseudo)vector and (pseudo)scalar.\\ 
In a more explicit way:
\begin{eqnarray}
\lambda_{c}^{(s)}S_{c}+\lambda_{u}^{(s)}S_{u}&=&
-b'_{1}+2b'_{2}+2b'_{4}+b'_{5}\nonumber \\ 
\lambda_{c}^{(s)}A_{c}+\lambda_{u}^{(s)}A_{u}&=&
-2b'_{7}-b'_{8}-4b'_{9}+b'_{10}\nonumber \\
\lambda_{c}^{(d)}S_{c}+\lambda_{u}^{(d)}S_{u}&=&
-b_{1}+2b_{2}+2b_{4}+b_{5}\nonumber \\ 
\lambda_{c}^{(d)}A_{c}+\lambda_{u}^{(d)}A_{u}&=&
-2b_{7}-b_{8}-4b_{9}+b_{10}.\nonumber \\
\end{eqnarray}
Using these decay modes we can, in principle, measure $\gamma$: 
\begin{itemize}
\item From these amplitudes we can obtain {\bf 10} equations 
(the process and the $CP$ conjugate process).
\item From $U$-spin theorem we obtain {\bf 2} constraints on these equations, namely \cite{uspin}:
\begin{eqnarray*}
|{\cal A}(B^{0}_{s} \rightarrow \rho^{+}\pi^{-})|^{2}-
|{\cal A}(\overline{B^{0}_{s}} \rightarrow \rho^{-}\pi^{+})|^{2} & = &
-|{\cal A}(B^{0} \rightarrow K^{*+}K^{-})|^{2}
-|{\cal A}(\overline{B^{0}} \rightarrow K^{*-}K^{+})|^{2}\\
|{\cal A}(B^{0}_{s} \rightarrow \rho^{-}\pi^{+})|^{2}-
|{\cal A}(\overline{B^{0}_{s}} \rightarrow \rho^{+}\pi^{-})|^{2} & = &
-|{\cal A}(B^{0} \rightarrow K^{*-}K^{+})|^{2}
-|{\cal A}(\overline{B^{0}} \rightarrow K^{*+}K^{-})|^{2}
\end{eqnarray*}
\item Totally we have {\bf 8} independent equations
\item We have {\bf 8} unknowns:
\begin{itemize}
\item $|S_{c}|$, $|S_{u}|$, $|A_{c}|$, $|A_{c}|$
\item 3 strong (relative) phases
\item 1 weak phase $\gamma$
\end{itemize}
\end{itemize}
Therefore we have enough equations to determine $\gamma$.\\
It should be noted that apart from the technical difficulty of solving 8 equations, the branching ratios of the various processes are expected to be small, since the spectator quark in each process does not appear in the final product. Using the graphical language this processes involve the diagrams $E$ and $PA$ which are expected to be small \cite{grlh}. Finally $SU(3)$ breaking effects should be taken into consideration.\\\\
To summarize, we have seen that the EWP contributions can be systematically related to the Tree contribution using the parameters $\delta_{EW}$ and $\bar{\delta}_{EW}$. A new constraint can be obtained using the decay mode $B^{0}\rightarrow K \pi$. 
Similar constraints can be obtained using $B^{0}\rightarrow \pi \pi$  
and $B^{0}_{s}\rightarrow K \pi$ but they are probably not valid.
Neubert-Rosner constraints cannot be obtained in 
$B \rightarrow VP$ without neglecting some parts of the amplitudes.
By neglecting some parts of the amplitudes we can obtain 
weaker constraints using $B^{0}\rightarrow \rho K$ or
$B^{0}\rightarrow K^{*} \pi$. Finally, using $B^{0}_{s} \rightarrow \rho\pi$ and $B^{0} \rightarrow K^{*\pm}K^{\mp}$ we can, in principle, measure $\gamma$ with some reservations.

\section*{\large \bf Acknowledgments}

I would like to thank Michael Gronau for useful discussions and Amnon Harel, Frank Krauss and Dan Pirjol for their helpful remarks.\\

\newpage
\appendix
\section{Algebraic Analysis of \mbox{\boldmath$B$} Decays}
\subsection{Algebraic Analysis of \mbox{\boldmath$B \rightarrow PP $} Decays}
In the following tables we give the $SU(3)$  decomposition of the decay amplitudes of B mesons to two pseudo-scalars mesons. In computing the tables we have used equation (\ref{hcomp0}) for $\Delta S=0$ decays and equation (\ref{hcomp1}) for $\Delta S=1$ decays. In every column of the table a decay amplitude is written as the sum of the integer number times $a_{i}$ (for $\Delta S=0$) decays or $a_{i}'$ ( for $\Delta S=1$ decays). 
Where:\\

$
\begin{array}{ll}
a_{1}=(\lambda^{(d)}_{u}a_{u}+\lambda^{(d)}_{c}a_{c})\sqrt{\frac{1}{12}}\langle{1}\|\overline{3}\|3\rangle & a'_{1}=(\lambda^{(s)}_{u}a_{u}+\lambda^{(s)}_{c}a_{c})\sqrt{\frac{1}{12}}\langle{1}\|\overline{3}\|3\rangle \\
a_{2}=(\lambda^{(d)}_{u}a_{u}+\lambda^{(d)}_{c}a_{c})\sqrt{\frac{1}{15}}\langle{8}_{S}\|\overline{3}\|3\rangle & a'_{2}=(\lambda^{(s)}_{u}a_{u}+\lambda^{(s)}_{c}a_{c})\sqrt{\frac{1}{15}}\langle{8}_{S}\|\overline{3}\|3\rangle \\
a_{3}=-\frac{1}{2}|\lambda^{(d)}_{u}|(c_{2}-c_{1}) 
(e^{i\gamma}+\bar{\delta}_{EW})\sqrt{\frac{1}{5}}\langle{8}_{S}\|6\|3\rangle & a'_{3}=-\frac{1}{2}|\lambda^{(s)}_{u}|(c_{2}-c_{1})
(e^{i\gamma}+\delta_{EW})\sqrt{\frac{1}{5}}\langle{8}_{S}\|6\|3\rangle\\ 
a_{4}=-\frac{1}{2}|\lambda^{(d)}_{u}|(c_{2}+c_{1})
(e^{i\gamma}-\bar{\delta}_{EW})\sqrt{\frac{1}{50}}\langle{8}_{S}\|\overline{15}\|3\rangle & a'_{4}=-\frac{1}{2}|\lambda^{(s)}_{u}|(c_{2}+c_{1})
(e^{i\gamma}-\delta_{EW})\sqrt{\frac{1}{50}}\langle{8}_{S}\|\overline{15}\|3\rangle\\    
a_{5}=-\frac{1}{2}|\lambda^{(d)}_{u}|(c_{2}+c_{1})
(e^{i\gamma}-\bar{\delta}_{EW})\sqrt{\frac{1}{200}}\langle{27}\|\overline{15}\|3\rangle &a'_{5}=-\frac{1}{2}|\lambda^{(s)}_{u}|(c_{2}+c_{1})
(e^{i\gamma}-\delta_{EW})\sqrt{\frac{1}{200}}\langle{27}\|\overline{15}\|3\rangle.
\end{array}$\\

As an example we take ${\cal A}(B^{+}\rightarrow\pi^{0}\pi^{+})$. From the table we find that : \\

$\sqrt{2}{\cal A}(B^{+}\rightarrow\pi^{0}\pi^{+})=+20a_{5}=
20\cdot(-\frac{1}{2}|\lambda^{(d)}_{u}|(c_{2}+c_{1})
(e^{i\gamma}-\bar{\delta}_{EW})\sqrt{\frac{1}{200}}\langle{27}\|\overline{15}\|3\rangle)$. \\

All the decays which include $\pi^{0}$ in the final state were written as $\sqrt{2}{\cal A}$, in order to simplify the notation.

\begin{table}[h]
\begin{center}
\begin{tabular}{|c|cccc|} \hline  
&$B^{0}_{s}\rightarrow K^{-}\pi^{+}$ & $\sqrt{2}(B^{0}_{s}\rightarrow \overline{K^{0}}\pi^{0}) $ &$B^{+}\rightarrow \overline{K^{0}} K^{+} $ & $\sqrt{2}(B^{+}\rightarrow \pi^{0} \pi^{+})$ \\ \hline 
$ a_{1}$&$0$&$0$&$0$&$0$\\ 
$a_{2}$&$-3$&$-3$&$-3$&$0$\\ 
$a_{3}$&$+1$&$+1$&$-1$&$0$\\ 
$a_{4}$&$-1$&$-1$&$+3$&$0$\\ 
$a_{5}$&$-8$&$+12$&$+4$&$+20$\\ 
\hline 
\hline
& $B^{0}\rightarrow K^{-} K^{+}$&$B^{0}\rightarrow \overline{K^{0}} K^{0}$&$B^{0}\rightarrow \pi^{-} \pi^{+}$&$\sqrt{2}(B^{0}\rightarrow \pi^{0} \pi^{0})$ \\ \hline 
$ a_{1}$&$-1$&$-1$&$-1$&$-1$\\ 
$a_{2}$&$+2$&$-1$&$-1$&$-1$\\ 
$a_{3}$&$0$&$-1$&$+1$&$+1$\\ 
$a_{4}$&$+2$&$-3$&$+1$&$+1$\\ 
$a_{5}$&$+1$&$+1$&$-7$&$+13$\\ 
\hline 
\end{tabular}\\
\caption{SU(3) decomposition of $B \rightarrow PP $ Decays: $\Delta S=0$}
\end{center}
\label{tabdsvp}
\end{table}


\begin{table}[t!!]
\begin{center}
\begin{tabular}{|c|cccc|} \hline  
&$B^{0}\rightarrow K^{+}\pi^{-}$ & $\sqrt{2}(B^{0}\rightarrow K^{0}\pi^{0}) $ &$\sqrt{2}(B^{+}\rightarrow K^{+}\pi^{0})$&$B^{+}\rightarrow K^{0} \pi^{+}$ \\ \hline 
$a'_{1}$&$0$&$0$&$0$&$0$\\ 
$a'_{2}$&$-3$&$-3$&$+3$&$-3$\\ 
$a'_{3}$&$+1$&$+1$&$+1$&$-1$\\ 
$a'_{4}$&$-1$&$-1$&$-3$&$+3$\\ 
$a'_{5}$&$-8$&$+12$&$+16$&$+4$\\ 
\hline 
\hline
& $B^{0}_{s}\rightarrow K^{-} K^{+}$&$B^{0}_{s}\rightarrow \overline{K^{0}} K^{0}$&$B^{0}_{s}\rightarrow \pi^{-} \pi^{+}$&$\sqrt{2}(B^{0}_{s}\rightarrow \pi^{0} \pi^{0})$ \\ \hline 
$a'_{1}$&$-1$&$-1$&$-1$&$-1$\\ 
$a'_{2}$&$-1$&$-1$&$+2$&$+2$\\ 
$a'_{3}$&$+1$&$-1$&$0$&$0$\\ 
$a'_{4}$&$+1$&$-3$&$+2$&$+2$\\ 
$a'_{5}$&$-7$&$+1$&$+1$&$+1$\\ 
\hline 
\end{tabular}\\
\caption{SU(3) decomposition of $B \rightarrow PP $ Decays: $\Delta S=1$}
\end{center}
\end{table}
\clearpage
\subsection{Algebraic Analysis of \mbox{\boldmath$B \rightarrow VP $} Decays}
In the following tables we give the $SU(3)$  decomposition of the decay amplitudes of $B$ mesons to pairs of mesons: the first being a vector meson and the second a pseudo-scalar. In computing the tables we have used  equation (\ref{hcomp0}) for $\Delta S=0$ decays and equation (\ref{hcomp1}) for $\Delta S=1$ decays. In every column of the table a decay amplitude is written as the sum of the integer number times $b_{i}$ (for $\Delta S=0$) decays or $b_{i}'$ ( for $\Delta S=1$ decays). Where:
\\

$
\begin{array}{ll}
b_{1}=(\lambda^{(d)}_{u}a_{u}+\lambda^{(d)}_{c}a_{c})\sqrt{\frac{1}{24}}\langle{1}\|\overline{3}\|3\rangle & b'_{1}=(\lambda^{(s)}_{u}a_{u}+\lambda^{(s)}_{c}a_{c})\sqrt{\frac{1}{24}}\langle{1}\|\overline{3}\|3\rangle \\
b_{2}=(\lambda^{(d)}_{u}a_{u}+\lambda^{(d)}_{c}a_{c})\sqrt{\frac{1}{30}}\langle{8}_{S}\|\overline{3}\|3\rangle & b'_{2}=(\lambda^{(s)}_{u}a_{u}+\lambda^{(s)}_{c}a_{c})\sqrt{\frac{1}{30}}\langle{8}_{S}\|\overline{3}\|3\rangle \\
b_{3}=-\frac{1}{2}|\lambda^{(d)}_{u}|(c_{2}-c_{1})
(e^{i\gamma}+\bar{\delta}_{EW})\sqrt{\frac{1}{10}}\langle{8}_{S}\|6\|3\rangle & b'_{3}=-\frac{1}{2}|\lambda^{(s)}_{u}|(c_{2}-c_{1})
(e^{i\gamma}+\delta_{EW})\sqrt{\frac{1}{10}}\langle{8}_{S}\|6\|3\rangle \\    
b_{4}=-\frac{1}{2}|\lambda^{(d)}_{u}|(c_{2}+c_{1})
(e^{i\gamma}-\bar{\delta}_{EW})\sqrt{\frac{1}{100}}\langle{8}_{S}\|\overline{15}\|3\rangle & b'_{4}=-\frac{1}{2}|\lambda^{(s)}_{u}|(c_{2}+c_{1})
(e^{i\gamma}-\delta_{EW})\sqrt{\frac{1}{100}}\langle{8}_{S}\|\overline{15}\|3\rangle \\  
b_{5}=-\frac{1}{2}|\lambda^{(d)}_{u}|(c_{2}+c_{1})
(e^{i\gamma}-\bar{\delta}_{EW})\sqrt{\frac{1}{400}}\langle{27}\|\overline{15}\|3\rangle & b'_{5}=-\frac{1}{2}|\lambda^{(s)}_{u}|(c_{2}+c_{1})
(e^{i\gamma}-\delta_{EW})\sqrt{\frac{1}{400}}\langle{27}\|\overline{15}\|3\rangle\\  
b_{6}=(\lambda^{(d)}_{u}a_{u}+\lambda^{(d)}_{c}a_{c})\sqrt{\frac{1}{6}}\langle{8}_{A}\|\overline{3}\|3\rangle & b'_{6}=(\lambda^{(s)}_{u}a_{u}+\lambda^{(s)}_{c}a_{c})\sqrt{\frac{1}{6}}\langle{8}_{A}\|\overline{3}\|3\rangle \\
b_{7}=-\frac{1}{2}|\lambda^{(d)}_{u}|(c_{2}-c_{1})
(e^{i\gamma}+\bar{\delta}_{EW})\sqrt{\frac{1}{18}}\langle{8}_{A}\|6\|3\rangle & b'_{7}=-\frac{1}{2}|\lambda^{(s)}_{u}|(c_{2}-c_{1})
(e^{i\gamma}+\delta_{EW})\sqrt{\frac{1}{18}}\langle{8}_{A}\|6\|3\rangle \\
b_{8}=-\frac{1}{2}|\lambda^{(d)}_{u}|(c_{2}-c_{1})
(e^{i\gamma}+\bar{\delta}_{EW})\sqrt{\frac{1}{36}}\langle{10}\|6\|3\rangle & b'_{8}=-\frac{1}{2}|\lambda^{(s)}_{u}|(c_{2}-c_{1})
(e^{i\gamma}+\delta_{EW})\sqrt{\frac{1}{36}}\langle{10}\|6\|3\rangle\\  
b_{9}=-\frac{1}{2}|\lambda^{(d)}_{u}|(c_{2}+c_{1})
(e^{i\gamma}-\bar{\delta}_{EW})\sqrt{\frac{1}{180}}\langle{8}_{A}\|\overline{15}\|3\rangle & b'_{9}=-\frac{1}{2}|\lambda^{(s)}_{u}|(c_{2}+c_{1})
(e^{i\gamma}-\delta_{EW})\sqrt{\frac{1}{180}}\langle{8}_{A}\|\overline{15}\|3\rangle \\    
b_{10}=-\frac{1}{2}|\lambda^{(d)}_{u}|(c_{2}+c_{1})
(e^{i\gamma}-\bar{\delta}_{EW})\sqrt{\frac{1}{36}}\langle\overline{10}\|\overline{15}\|3\rangle & b'_{10}=-\frac{1}{2}|\lambda^{(s)}_{u}|(c_{2}+c_{1})
(e^{i\gamma}-\delta_{EW})\sqrt{\frac{1}{36}}\langle\overline{10}\|\overline{15}\|3\rangle 
\end{array}$\\\\

 It is easy to see that there is a connection between the coefficients of $B\rightarrow PP$  and $B\rightarrow VP$, namely,   $\sqrt{2}b_{i}=a_{i}$ and $\sqrt{2}b_{i}'=a_{i}'$ (where $1\leq i \leq 5 $). This connection is a result of the symmetrization of the final state in $B\rightarrow PP$ decays.\\  
In some of the decays which include $\pi^{0}$ or $\rho^{0}$ in the final state were written as $\sqrt{2}{\cal A}$, in order to simplify the notation. 

\begin{table}[h]
\begin{center}
\begin{tabular}{|c|cccc|} \hline  
&$B^{+}\rightarrow\overline{K^{*0}}K^{+}$&$\sqrt{2}(B^{+}\rightarrow\rho^{0}\pi^{+})$&$B^{+}\rightarrow K^{*+}\overline{K^{0}}$&$\sqrt{2}(B^{+}\rightarrow\rho^{+}\pi^{0})$\\ \hline   
$b_{1}$&$0$&$0$&$0$&$0$\\ 
$b_{2}$&$-3$&$0$&$-3$&$0$\\ 
$b_{3}$&$-1$&$0$&$-1$&$0$\\ 
$b_{4}$&$+3$&$0$&$+3$&$0$\\ 
$b_{5}$&$+4$&$+20$&$+4$&$+20$\\ \hline
$b_{6}$&$-1$&$-2$&$+1$&$+2$\\ 
$b_{7}$&$-1$&$-2$&$+1$&$+2$\\ 
$b_{8}$&$-2$&$+2$&$+2$&$-2$\\ 
$b_{9}$&$+3$&$+6$&$-3$&$-6$\\ 
$b_{10}$&$0$&$0$&$0$&$0$\\ 
\hline 
\end{tabular}\\
\caption{SU(3) decomposition of $B^{+}\rightarrow VP$ : $\Delta S=0$ }
\end{center}
\end{table}

\begin{table}[h]
\begin{center}
\begin{tabular}{|c|cccc|} \hline  
&$\sqrt{2}(B^{+}\rightarrow\rho^{0} K^{+})$&$B^{+}\rightarrow\rho^{+} K^{0}$&$
\sqrt{2}(B^{+}\rightarrow K^{*+} \pi^{0})$&$B^{+}\rightarrow K^{*0} \pi^{+}$
\\ \hline 
$b'_{1}$&$0$&$0$&$0$&$0$\\ 
$b'_{2}$&$+3$&$-3$&$+3$&$-3$\\ 
$b'_{3}$&$+1$&$-1$&$+1$&$-1$\\ 
$b'_{4}$&$-3$&$+3$&$-3$&$+3$\\ 
$b'_{5}$&$+16$&$+4$&$+16$&$+4$\\ \hline
$b'_{6}$&$-1$&$+1$&$+1$&$-1$\\ 
$b'_{7}$&$-1$&$+1$&$+1$&$-1$\\ 
$b'_{8}$&$+4$&$+2$&$-4$&$-2$\\ 
$b'_{9}$&$+3$&$-3$&$-3$&$+3$\\ 
$b'_{10}$&$0$&$0$&$0$&$0$\\ 
\hline 
\end{tabular}\\
\caption{SU(3) decomposition of $B^{+}\rightarrow VP$ : $\Delta S=1$ }
\end{center}
\end{table}

\begin{table}[h]
\begin{center}
\begin{tabular}{|c|cccc|} \hline  
&$B^{0}_{s}\rightarrow K^{*-}\pi^{+}$&$\sqrt{2}(B^{0}_{s}\rightarrow\overline{K^{*0}}\pi^{0})$&$B^{0}_{s}\rightarrow\rho^{+}K^{-} $&$\sqrt{2}(B^{0}_{s}\rightarrow\rho^{0}\overline{K^{0}})$\\ \hline   
$b_{1}$&$0$&$0$&$0$&$0$\\ 
$b_{2}$&$-3$&$-3$&$-3$&$-3$\\ 
$b_{3}$&$+1$&$+1$&$+1$&$+1$\\ 
$b_{4}$&$-1$&$-1$&$-1$&$-1$\\ 
$b_{5}$&$-8$&$+12$&$-8$&$+12$\\ \hline
$b_{6}$&$+1$&$+1$&$-1$&$-1$\\ 
$b_{7}$&$-1$&$-1$&$+1$&$+1$\\
$b_{8}$&$-2$&$-2$&$+2$&$+2$\\ 
$b_{9}$&$+1$&$+1$&$-1$&$-1$\\ 
$b_{10}$&$+2$&$-4$&$-2$&$+4$\\ 

\hline 
\end{tabular}\\
\caption{SU(3) decomposition of $B^{0}_{s}\rightarrow VP$ : $\Delta S=0$}
\end{center}
\end{table}

\begin{table}[h]
\begin{center}
\begin{tabular}{|c|cccc|} \hline  
&$B^{0}\rightarrow\rho^{-} K^{+}$&$\sqrt{2}(B^{0}\rightarrow\rho^{0} K^{0})$&$
B^{0}\rightarrow K^{*+} \pi^{-}$&$\sqrt{2}(B^{0}\rightarrow K^{*0} \pi^{0})$
\\ \hline 
$b'_{1}$&$0$&$0$&$0$&$0$\\ 
$b'_{2}$&$-3$&$-3$&$-3$&$-3$\\ 
$b'_{3}$&$+1$&$+1$&$+1$&$+1$\\ 
$b'_{4}$&$-1$&$-1$&$-1$&$-1$\\ 
$b'_{5}$&$-8$&$+12$&$-8$&$+12$\\ \hline
$b'_{6}$&$+1$&$+1$&$-1$&$-1$\\ 
$b'_{7}$&$-1$&$-1$&$+1$&$+1$\\ 
$b'_{8}$&$-2$&$+4$&$+2$&$-4$\\ 
$b'_{9}$&$+1$&$+1$&$-1$&$-1$\\ 
$b'_{10}$&$+2$&$+2$&$-2$&$-2$\\ 
\hline 
\end{tabular}\\
\caption{SU(3) decomposition of $B^{0}\rightarrow VP$ : $\Delta S=1$ }
\end{center}
\end{table}

\begin{table}[h]
\begin{center}
\begin{tabular}{|c|ccccccc|} \hline 

&$K^{*-}K^{+}           $& 
$\overline{K^{*0}}K^{0}$& 
$\rho^{-}\pi^{+}       $& 
$K^{*+}K^{-}           $& 
$K^{*0}\overline{K^{0}}$& 
$\rho^{+}\pi^{-}       $& 
$\rho^{0}\pi^{0}       $\\ \hline

$b_{1}$&$-1$&$-1$&$-1$&$-1$&$-1$&$-1$&$-1$\\ 
$b_{2}$&$+2$&$-1$&$-1$&$+2$&$-1$&$-1$&$-1$\\ 
$b_{3}$&$0$&$-1$&$+1$&$0$&$-1$&$+1$&$+1$\\ 
$b_{4}$&$+2$&$-3$&$+1$&$+2$&$-3$&$+1$&$+1$\\ 
$b_{5}$&$+1$&$+1$&$-7$&$+1$&$+1$&$-7$&$13$\\ \hline
$b_{6}$&$0$&$-1$&$+1$&$0$&$+1$&$-1$&$0$\\ 
$b_{7}$&$+2$&$+1$&$+1$&$-2$&$-1$&$-1$&$0$\\
$b_{8}$&$+1$&$-1$&$-1$&$-1$&$+1$&$+1$&$0$\\ 
$b_{9}$&$+4$&$-1$&$+5$&$-4$&$+1$&$-5$&$0$\\ 
$b_{10}$&$-1$&$+1$&$+1$&$+1$&$-1$&$-1$&$0$\\ 

\hline 
\end{tabular}\\
\caption{SU(3) decomposition of $B^{0}\rightarrow VP$ : $\Delta S=0$}
\end{center}
\end{table}

\begin{table}[h]
\begin{center}
\begin{tabular}{|c|ccccccc|} \hline 

&$K^{*-}K^{+}          $& 
$\overline{K^{*0}}K^{0}$& 
$\rho^{-}\pi^{+}       $& 
$K^{*+}K^{-}           $& 
$K^{*0}\overline{K^{0}}$& 
$\rho^{+}\pi^{-}       $& 
$\rho^{0}\pi^{0}       $\\ \hline 

$b'_{1}$&$-1$&$-1$&$-1$&$-1$&$-1$&$-1$&$-1$\\ 
$b'_{2}$&$-1$&$-1$&$+2$&$-1$&$-1$&$+2$&$+2$\\ 
$b'_{3}$&$+1$&$-1$&$0$&$+1$&$-1$&$0$&$0$\\ 
$b'_{4}$&$+1$&$-3$&$+2$&$+1$&$-3$&$+2$&$+2$\\ 
$b'_{5}$&$-7$&$+1$&$+1$&$-7$&$+1$&$+1$&$+1$\\ \hline
$b'_{6}$&$+1$&$+1$&$0$&$-1$&$-1$&$0$&$0$\\ 
$b'_{7}$&$+1$&$-1$&$+2$&$-1$&$+1$&$-2$&$0$\\
$b'_{8}$&$-1$&$+1$&$+1$&$+1$&$-1$&$-1$&$0$\\ 
$b'_{9}$&$+5$&$+1$&$+4$&$-5$&$-1$&$-4$&$0$\\ 
$b'_{10}$&$+1$&$-1$&$-1$&$-1$&$+1$&$+1$&$0$\\ 

\hline 
\end{tabular}\\
\caption{SU(3) decomposition of $B^{0}_{s}\rightarrow VP$ : $\Delta S=1$}
\end{center}
\end{table}

\clearpage
\section{Explicit Expressions for \mbox{\boldmath$a_{u}$} and
 \mbox{\boldmath$a_{c}$}}

In section 2 we were able to write the effective Hamiltonian in a very compact way - equation (\ref{hcomp1}) and (\ref{hcomp0}). These expressions contain in them the coefficients $a_{u}$ and $a_{c}$. Eventually $a_{u}$ and $a_{c}$ appear in the decomposition of the amplitudes.\\
 Let us denote by: $\langle\cdot\|$
the ``kets'' $\langle 1\|$, $\langle{8}_{S}\|$ and $\langle{8}_{A}\|$.\\
 Among the $\overline{3}$ operators we have:
\begin{itemize}
\item operators with helicity structure $V-A,V-A$:\\\\ 
$\overline{3}^{(s)}$, $\overline{3}^{(a)}$ and $\overline{3}$ (the charming penguin operator).
\item operators with helicity structure $V-A,V+A$:\\\\  
$\overline{3}^{(s)}_{\pm}$, $\overline{3}^{(a)}_{\pm}$ and $\overline{3}_{\pm}$
\end{itemize}
therefore we can write:
\begin{eqnarray} 
a_{u}\langle\cdot\|\overline{3}\|3\rangle&=&
\langle\cdot\|\overline{3}^{(s)}\|3\rangle\left\{
\frac{1}{\sqrt{8}}\left[c_{2}+c_{1}+\frac{3}{2}\left(c_{9}+c_{10}\right)\right]
+\sqrt{2}\left[c_{3}+c_{4}-\frac{1}{2}\left(c_{9}+c_{10}\right) \right]\right\}
\nonumber \\
&+&\langle\cdot\|\overline{3}^{(a)}\|3\rangle\left\{
-\frac{1}{2}\left[c_{2}-c_{1}-\frac{3}{2}\left(c_{9}+c_{10}\right)\right]
+\left[c_{3}-c_{4}-\frac{1}{2}\left(c_{9}-c_{10}\right) \right]\right\}\nonumber \\
&+&\langle\cdot\|\overline{3}\|3\rangle\left\{c_{3}+c_{4}+c_{9}+c_{10}\right\}\nonumber \\
&+&\langle\cdot\|\overline{3}^{(s)}_{\pm}\|3\rangle\left\{\sqrt{2}\left(c_{5}+c_{6}\right)\right\}\nonumber \\
&+&\langle\cdot\|\overline{3}^{(a)}_{\pm}\|3\rangle\left\{c_{5}-c_{6}\right\}\nonumber \\
&+&\langle\cdot\|\overline{3}_{\pm}\|3\rangle\left\{c_{5}+c_{6}\right\} \\\nonumber \\
a_{c}\langle\cdot\|\overline{3}\|3\rangle&=&
\langle\cdot\|\overline{3}^{(s)}\|3\rangle\left\{
\frac{1}{\sqrt{8}}\left[\frac{3}{2}\left(c_{9}+c_{10}\right)\right]
+\sqrt{2}\left[c_{3}+c_{4}-\frac{1}{2}\left(c_{9}+c_{10}\right) \right]\right\}
\nonumber \\
&+&\langle\cdot\|\overline{3}^{(a)}\|3\rangle\left\{
+\frac{1}{2}\left[\frac{3}{2}\left(c_{9}+c_{10}\right)\right]
+\left[c_{3}-c_{4}-\frac{1}{2}\left(c_{9}-c_{10}\right) \right]\right\}\nonumber \\
&+&\langle\cdot\|\overline{3}\|3\rangle\left\{c_{3}+c_{4}+c_{9}+c_{10}\right\}\nonumber \\
&+&\langle\cdot\|\overline{3}^{(s)}_{\pm}\|3\rangle\left\{\sqrt{2}\left(c_{5}+c_{6}\right)\right\}\nonumber \\
&+&\langle\cdot\|\overline{3}^{(a)}_{\pm}\|3\rangle\left\{c_{5}-c_{6}\right\}\nonumber \\
&+&\langle\cdot\|\overline{3}_{\pm}\|3\rangle\left\{c_{5}+c_{6}\right\}
\end{eqnarray}

\end{document}